\documentclass{ab}

\usepackage{graphicx}
\usepackage{latexsym}

\def\degr{\hbox{$^\circ$}}

\begin{document}

\title{Simulation of a Zenith-Field Sky Survey  on RATAN-600 }

\author{E.K.~Majorova$^{1}$}
\institute{Special Astrophysical Observatory Russian Academy of Sciences,
N. Arkhyz, KChR, 369167, Russia}

\offprints{E.K.Majorova, \email{len@sao.ru}}

\titlerunning{Simulation of a Zenith-Field Sky Survey}

\authorrunning{Majorova }
 
\date{Received: November 1, 2007/Revised:November 29, 2007}

\abstract{
In this paper we simulate a deep multi-frequency zenith-field sky
survey on RATAN-600 (the RZF survey). In our simulations we use
the 1.4-GHz sky images obtained in the  NVSS survey. We convolved NVSS
images with the two-dimensional power beam pattern of RATAN-600
and obtain simulated 24-hour scans of sky transits at all
wavelengths of the RZF survey. For the 7.6-cm wavelength we
analyze the effect of the image area size on the results of the
simulation. We estimate the accuracy of the determination of
source fluxes on simulated scans and derive the distributions of
the spectral indices of the sources. We use the simulated scans to
clean real records of the RZF survey at 7.6~cm. The standard
error of the residual noise at this wavelength is about 1~mJy.
}
\maketitle

\section{INTRODUCTION}

Sky surveys are among the main sources of information about
cosmic objects. The first deep sky survey on RATAN-600 was
carried out in 1980--1982 during the ``Cold'' experiment at
the declination of $\delta \sim 5\degr$  [\cite{p2:Majorova_n}]. The
use of the best radiometer in terms of fluctuation sensitivity
made it possible to obtain new constraints on the background
radiation of the Universe and prepare a catalog of radio sources
with a detection threshold of $\sim10$~mJy (RC catalog) at 7.6~cm
 [\cite{p3:Majorova_n,p4:Majorova_n}].

During the period from 1979 through 1987 the multi-frequency
``Zelenchuk Survey'' in the declination interval
$\delta=0\degr\div14\degr$ [\cite{a1:Majorova_n,a2:Majorova_n}],
and during the period from 1987 through 1989, a polar survey
[\cite{a3:Majorova_n}], were carried out on the Southern sector
with a periscopic reflector. The threshold of 3.9-GHz source
detection flux was equal to 50 and 14~mJy for the ``Zelenchuk'' and
polar surveys, respectively. In 1988 the sky area $\delta =
47\degr06'45''\div47\degr07'45''$, $8^h\le\alpha< 14^h$ was
observed with the whole ring aperture of RATAN-600 with a
detection threshold of  $\sim 15$~mJy [\cite{v1:Majorova_n}].

The next step in conducting surveys on RATAN-600 was taken in the late 1990ies. The upgrade of
the radio telescope and substantial improvement of the radiometer sensitivity made it possible
to perform a deeper survey at a higher quality level.

The RZF survey (RATAN-600 Zenith Field) was performed during the period from 1998 through 2003 on
the Northern sector at wavelengths $1.0 \div 55$~cm at declination of 3ó84. At the main
wavelength of observations ($\lambda7.6$~cm) the half-power survey width was
$\delta_{2000.0} = 41\degr30'42''\pm2'$, $0^h\le\alpha_{2000.0}< 24^h$.
As a result of the survey, a catalog of radio sources at 7.6~cm (the RZF catalog) was
compiled [\cite{p1:Majorova_n}]. The minimum flux of the radio sources of the RZF catalog was
close to
about $2.5$~mJy, which is comparable to the threshold flux of the NVSS catalog [\cite{co1:Majorova_n}].
Starting from 2001, eight sky bands were observed in addition to the central band. The additional
bands were shifted by  $\Delta \delta = \pm12'n$ (n=1,~2,~3,~4) with respect to the central band.
Note that the density of the survey bands in declination was increased by a factor of two since
2006, and the interval between the neighboring bands became equal to $6'$.

A distinctive feature of the sky surveys performed with RATAN-600
compared, e.g., to VLA surveys (NVSS [\cite{co1:Majorova_n}], FIRST
[\cite{fr:Majorova_n}]), is that at a given declination  we observe
sources located within a certain interval (band) of declinations.
The angular size of this band is  determined by the size of the
vertical power beam (PB) of the radio telescope. The only
exception is the sky survey performed in 1988 in the mode of
whole circular aperture of the antenna with a ``pencil-shaped''
beam of the radio telescope at the half-power level. In all other
surveys one of the sectors of RATAN-600 was used (1/4 of the ring
aperture). In this mode of operation the power beam pattern
of the radio telescope differs substantially from that of a
paraboloid.

\begin{figure}[tbp]
\centerline{
\hbox{
\includegraphics[width=0.38\textwidth,clip,]{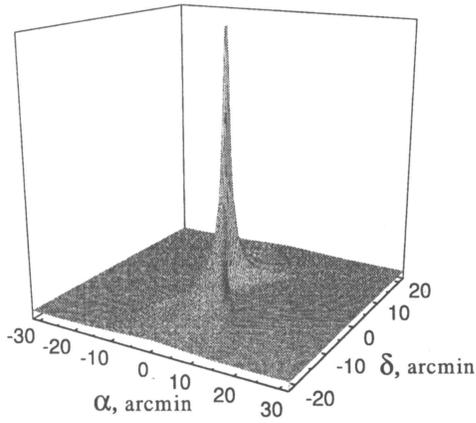}
}
}
\caption{Power beam pattern of RATAN-600 computed for the wavelength and elevation
 of 7.6~cm and
$H=88\degr42''$, respectively.}
\label{fig1:Majorova_n}
\end{figure}

A distinctive feature of the power beam  of RATAN-600
operating in a single-sector mode is its large extension in the
vertical plane and the extended ``fan-shaped'' background at large
elevations, whose intensity decreases as  1/y in the vertical
plane [\cite{m1:Majorova_n,m2:Majorova_n}].
Figure~~\ref{fig1:Majorova_n} gives an example of a  PB
computed for the elevation angle $H=88\degr42''$. Such a
structure of the power beam  allows, on the one hand,
sources to be observed within a certain declination band and, on
the other hand, makes the sources difficult to distinguish and
thus results in extended structures appearing on the records of
sky transits across the fixed PB of the telescope (the so-called
sky scans). This is because the magnitude and shape of the signal
from an individual source depend substantially on the distance of
the source from the central section of the power beam pattern.

\begin{figure}[tbp]
\centerline{
\hbox{
\includegraphics[width=0.35\textwidth,clip,]{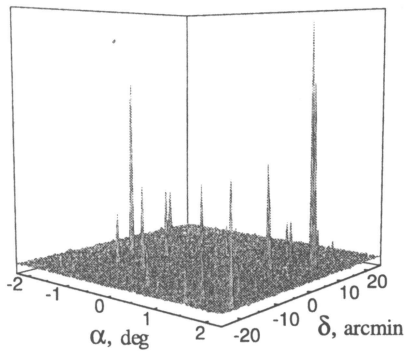}
}
}
\caption{Fragment of the $4\degr\times 4\degr$ NVSS image. The size in declination is trimmed
($\delta=\pm20'$).}
\label{fig2:Majorova_n}
\end{figure}

In view of the aforesaid, simulation of sky surveys conducted
with RATAN-600 is of considerable practical interest. Such
simulations are necessary both for studies of extended cosmic
objects in order to estimate the contribution of point sources to
their flux, and for extraction of numerous weak sources
against noise background, thereby reducing the effect of
``confusion''.

In this paper we attempt to simulate the  RZF sky survey in its
central section $\delta_{o} = 41\degr30'42''$ using the 1.4-GHz
sky images (charts) obtained as a result of the  NVSS (NRAO VLA
Sky Survey) [\cite{co1:Majorova_n}].

\section {TECHNIQUE OF SIMULATION OF THE ZENITH-FIELD SURVEY}

Bursov and Majorova [\cite{b1:Majorova_n}] were the first to
simulate the zenith-field sky survey (RZF-survey). To this end,
the above authors used the data of the NVSS catalog, from which
they adopted the principal parameters of the sources. The program
package for computing the two-dimensional PB of RATAN-600
[\cite{m1:Majorova_n}] was used to simulate the curves of the
transit of sources across various horizontal sections of the
power beam pattern. These computations assumed all sources to be
points, i.e., their angular sizes were assumed to be much smaller
than the halfwidth of the power beam  (HPBW). In this
case, the response of the antenna to the transit of a source with
declination  $\delta_{ist}$ across the power beam  of the
radio telescope  is identical to the horizontal section of the
power beam  shifted by $\Delta \delta$ from the central
section and multiplied by the 7.6-cm flux of the source in
question. Here $\Delta\delta=\delta_{ist}-\delta_{0}$, where
$\delta_{0}$ is the declination of the central section of the RZF
survey, which passed through 3ó84.

We computed the transit curves for most of the sources of the
NVSS catalog, whose declinations fall within the
$\Delta\delta=41\degr30'42''\pm 30'$ band. To compute the 7.6-cm
fluxes of the sources, we used their 21-cm fluxes adopted from
the NVSS catalog and the spectral indices of the corresponding
sources. For the sources with unavailable spectral indices we set
the spectral index equal to $\gamma=-0.8$, which corresponds to a
normal nonthermal spectrum ($S_{\nu}\sim\nu^{\gamma}$). We then
time shifted the curves in accordance with the right ascension
($\alpha$) of each particular source, and summed them up. The
resulting one-dimensional scan of the diurnal transit of the sky
band across the fixed PB of the radio telescope provided a good
analog of real observations. Let us refer this model of the
RZF-survey as the model based on the NVSS catalog.

\begin{figure*}[tbp]
\centerline{
\hbox{
\includegraphics[angle=-90,width=0.6\textwidth,clip]{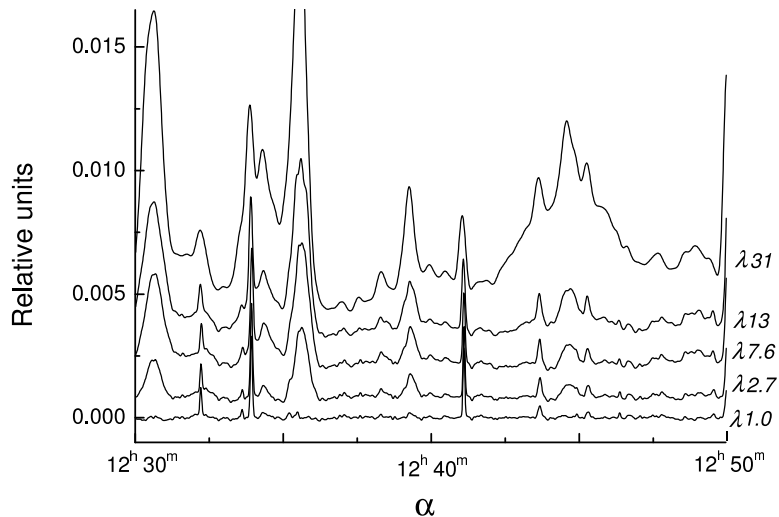}
} }
\caption{
Normalized simulated 20-minute scans of the
$\delta_{2000.0}=41\degr30'42''$, $12^{h}30^{m} \le
\alpha_{2000.0} \le 12^{h}50^{m}$ sky area at wavelengths 1.0,
2.7, 7.6, 13, and 31~cm obtained by convolving the PB of RATAN-600
with the $4\degr\times 4\degr$ areas of NVSS images. We
normalized the scans to the signal level from 3ó84.The curves are
shifted only along the vertical axis.} \label{fig3:Majorova_n}
\end{figure*}

In this paper we simulate the zenith-field survey by convolving the
two-dimensional PB of RATAN-600 with the VLA NVSS sky images with
a resolution of $\theta=45''$, which have the form of
$4\degr\times4\degr$ areas. The images are in the FITS format and
they are available from the NRAO site [\cite{na1:Majorova_n}]. Each
of the 2326 1024$\times$1034 images has a  name in the form
PHHMMSDD, where P is the name of the Stokes parameter (I, Q, or
U); îî and MM are the hours and minutes of right ascension of
the image center; S, the sign of the declination (P and M for +
and $-$, respectively), and DD is the declination of the image
center in degrees for the epoch of 2000 (J2000).

To simulate the RZF survey, we use about $20^m$-long intensity
images  Ihh00PDD, Ihh20PDD, and Ihh40PDD (where $hh = 00,
01,..., 23$) with the central declination of $\delta_{0} =
+40\degr$ for the simulation at $\lambda \le 7.6$~cm; $\delta_{0}
= +40\degr$ and $+44\degr$ for the simulation at $\lambda13$ and
31~cm, and $\delta_{0} = +36\degr$, $+40\degr$, and $+44\degr$for
the simulation at $\lambda49$~cm.

We compute the two-dimensional power beam pattern of RATAN-600 as described
by Majorova [\cite{m1:Majorova_n}]. We performed our computations for
the elevation of 3C84 ($H=88\degr42''$) in such declination
intervals that maximum values of the power beam  in the
most distant sections are equal to a fraction of percent of the
maximum of the diagram in its central section. The width of this
declination interval was equal to  $d\delta=\pm6'$, $d\delta =
\pm 37'15''$, and  $d\delta=\pm 4\degr53'$ $ for \lambda1.0$,
$\lambda7.6,$ and $\lambda49$, respectively. We compute the
power beam pattern of the radio telescope with a step equal to
the pixel size on the NVSS survey maps (15 arcsec).
Figure~\ref{fig1:Majorova_n} shows the two-dimensional power beam
computed for the wavelength of 7.6~cm and elevation of
$H=88\degr42''$. The size of the power beam  in the
figure is trimmed, $d\delta=\pm20'$.

We convolved the diagram with the part of the NVSS-survey images whose
declination size is equal to the vertical size of the computed PB
for the corresponding wavelength ($d\delta$). The center of NVSS
images was located at the declination of $\delta=41\degr30'45''$,
which virtually coincides with the declination of 3ó84. In
addition to the  IHHMMP40 images we used  IHHMMP44
(at wavelengths 7.6~cm <$\lambda$<49~cm) and  IHHMMP36 and
IHHMMP44 (at the 49-cm wavelength) images. Figure~\ref{fig2:Majorova_n}
shows an example of an NVSS image with a declination size of
$d\delta=\pm20'$.

We convert the images from the FITS format into binary matrices,
which we then convolved with two-dimensional PB. As a result of these
convolutions, we obtain 72 one-dimensional 20-minute sky scans at
each wavelength. Note that the declination size of NVSS images
(and that of the power beam) exceed the width of the
survey band at the wavelength considered by more than one order
of magnitude, thereby allowing better account to be taken of
strong distant sources falling within the PB of the radio
telescope. Further increase of image sizes in $\delta$ has
virtually no effect on the results of the simulation.

We normalize the resulting scans to the level of the 3C84 signal
and convert them into F-format files. F format is a modification
of FITS format used on RATAN-600 for representing the actual
records and for their subsequent reduction.

Figure~\ref{fig3:Majorova_n} shows the normalized simulated
20-minute scans of the  $\delta_{2000.0}=41\degr30'42''$,
$12^{h}30^{m} \le \alpha_{2000.0} \le 12^{h}50^{m}$  sky area  at
1.0, 2.7, 7.6, 13, and 31~cm wavelengths obtained using the
technique described above. Here and in subsequent figures scans
are shifted with respect to each other along the vertical axis
for better visualization.

It is evident from the curves presented that as we pass from
$\lambda1.0$~cm to longer wavelengths the halfwidths of the
sources in records increase and new sources and extended
structures appear. This is due to the increase of the size of the
PB with increasing wavelength, so that the power beam
begins to cover the sky areas increasingly distant from the
central section. Strong sources passing far from the central
section  in terms of $\delta$ would produce extended features
similar to the one we see in Fig.~\ref{fig3:Majorova_n} at
$\lambda13$ and 31~cm in the right-ascension interval $
12^{h}42^{m} <  \alpha < 12^{h}47^{m}$.

\begin{figure}[tbp]
\centerline{
\vbox{
\hbox{
\includegraphics[angle=-90,width=0.42\textwidth,clip]{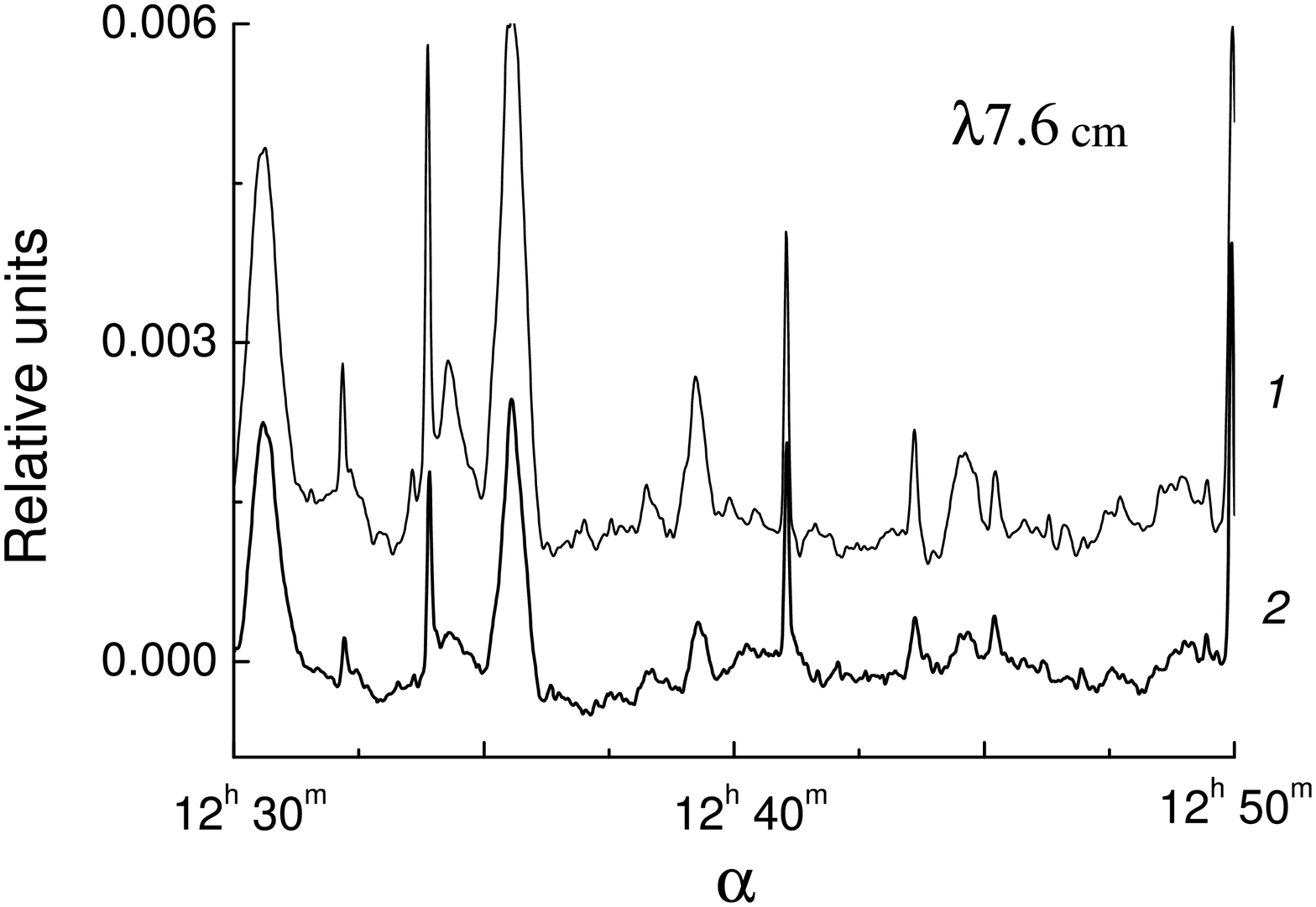}
}
\hbox{
\includegraphics[angle=-90,width=0.42\textwidth,clip]{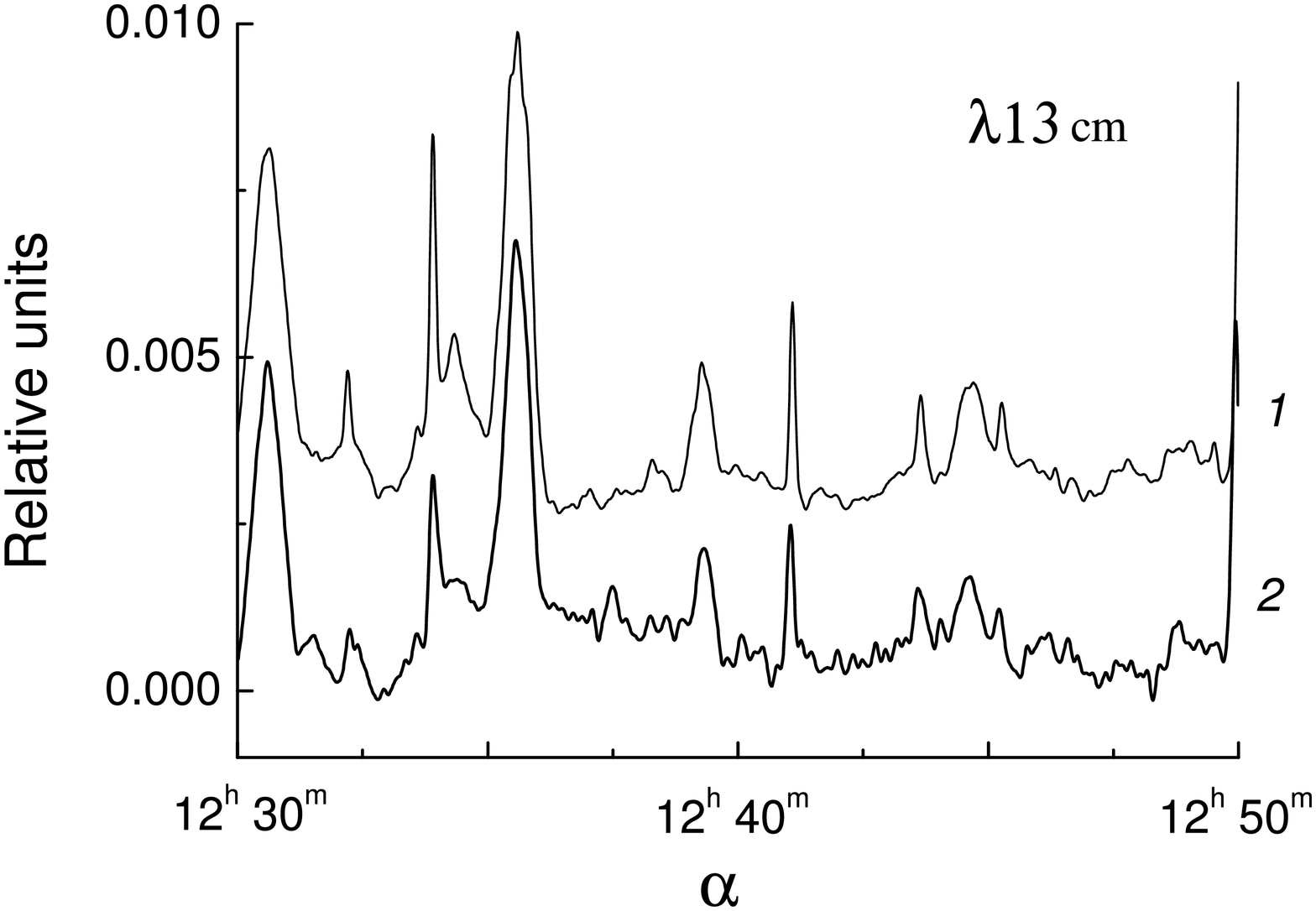}
}
\hbox{
\includegraphics[angle=-90,width=0.42\textwidth,clip]{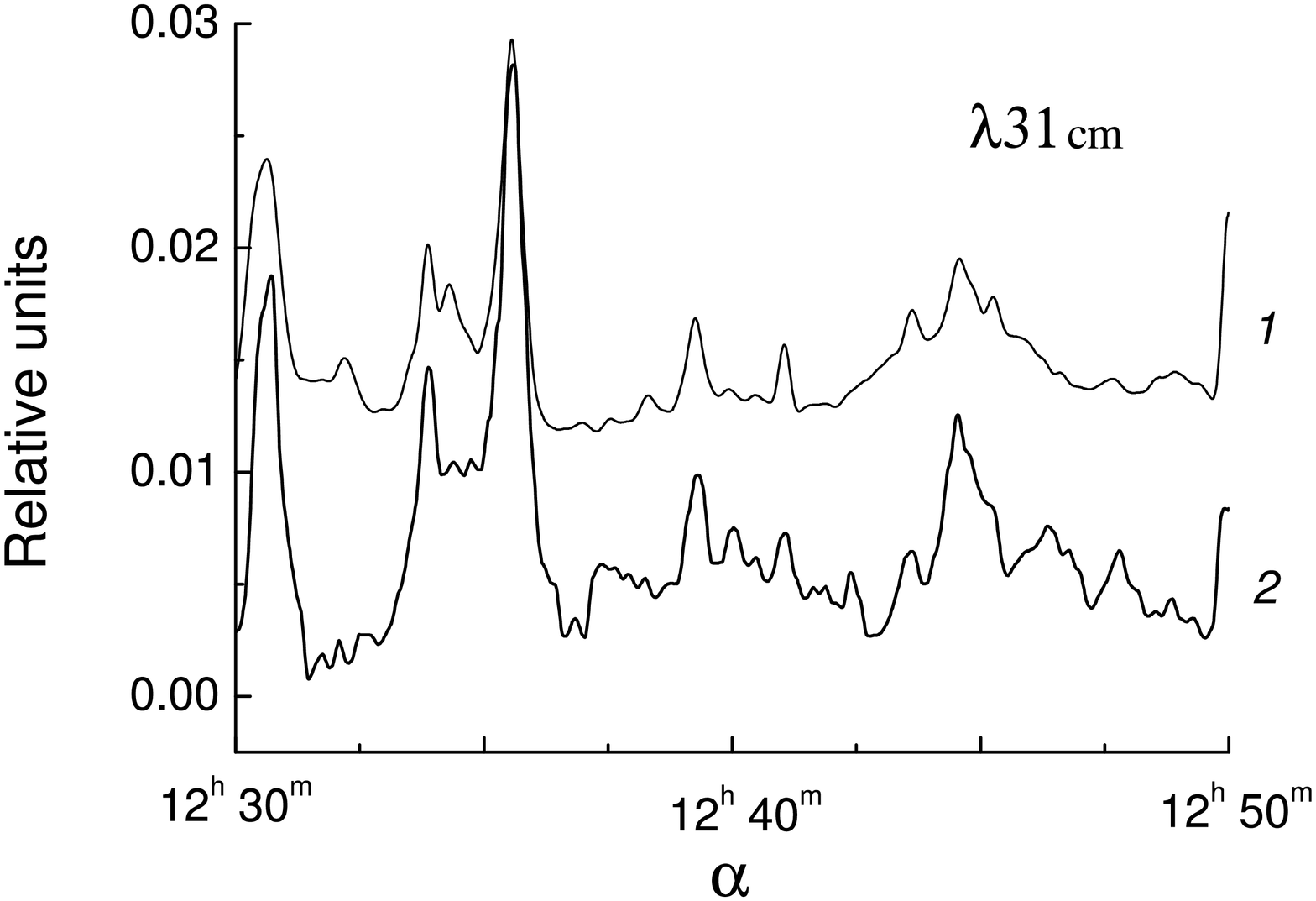}
} } }
\caption{
Normalized simulated scans at the wavelengths of 7.6, 13, and
31~cm obtained using NVSS images (curves \emph{1}), and
normalized real averaged records of the same sky areas obtained
within the framework of RZF survey (curves \emph{2}). The scans
are normalized to the level of the 3C84 signal. The curves are
shown shifted along the the vertical axis. } \label{fig4:Majorova_n}
\end{figure}

Figure~\ref{fig4:Majorova_n} shows normalized simulated scans at 7.6, 13, and 31~cm (curves
\emph{1}) and the corresponding normalized transit curves obtained within the framework of
the RZF survey (curves \emph{2}). The real records shown in the figures were kindly provided
by N.~N.~Bursov. They were obtained by averaging about 200 transits of the same sky area in the
central band of the survey during the period from 1998 through 2003 [\cite{b2:Majorova_n}]. The
simulated and real scans are normalized to the level of the signal from 3ó84.

The subsequent analysis is based on the results of the simulation
of the RZF survey at 7.6~cm. The sensitivity of the RZF survey is
maximum at this wavelength and it is close to the sensitivity
of the NVSS  survey.

\begin{figure}[tbp]
\centerline{
\vbox{
\hbox{
\includegraphics[angle=-90,width=0.47\textwidth,clip]{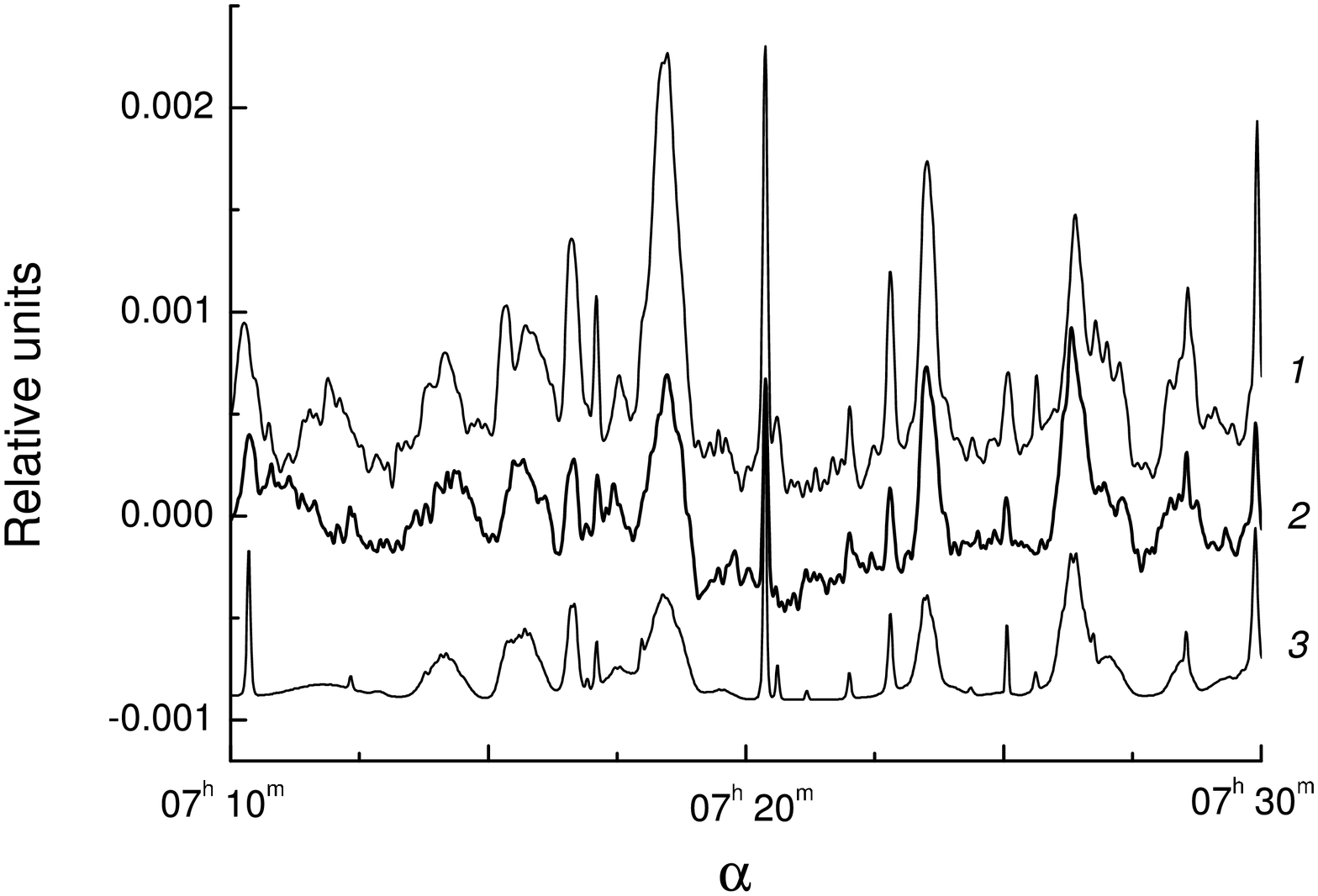}
}
\hbox{
\includegraphics[angle=-90,width=0.47\textwidth,clip]{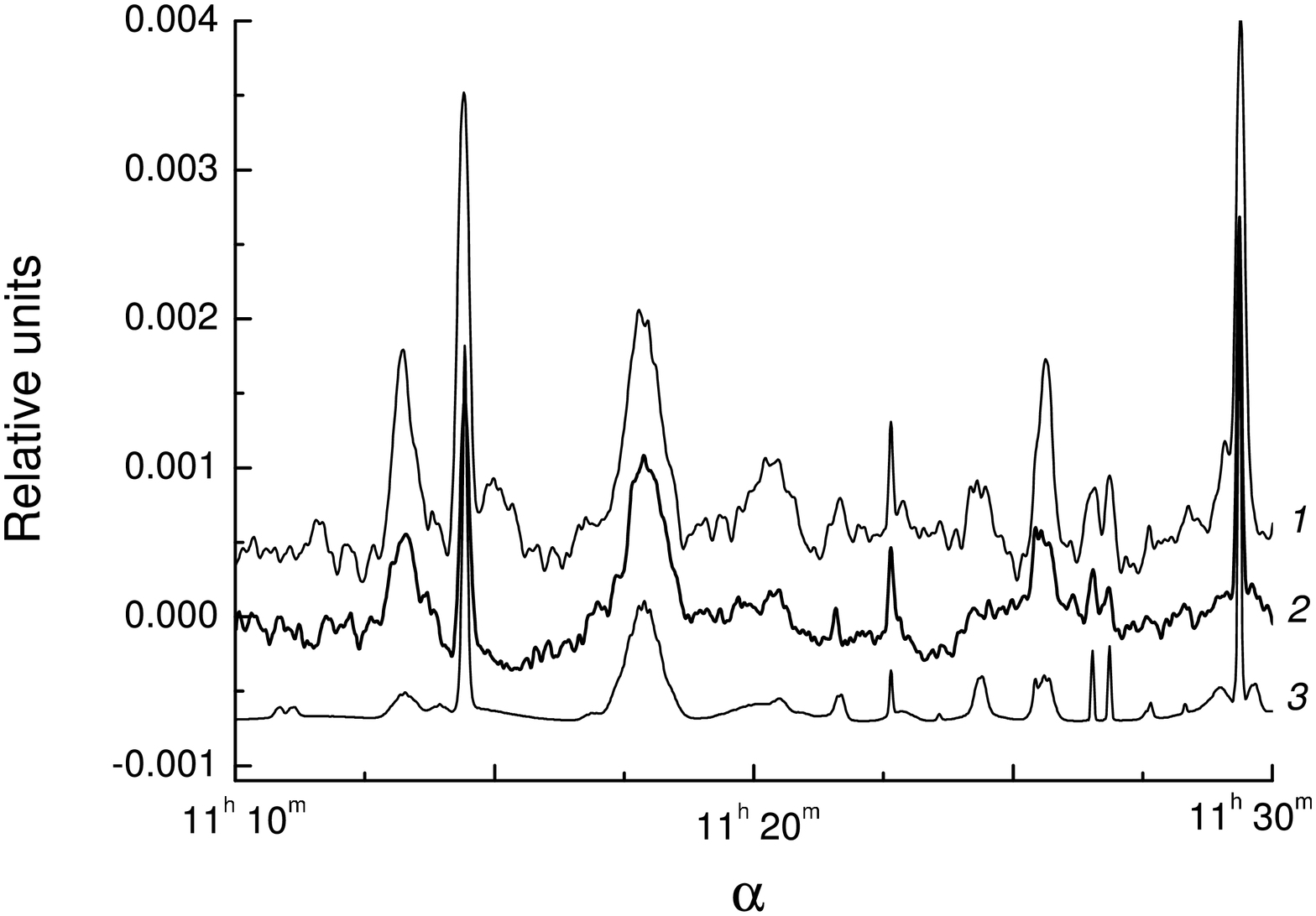}
}
\hbox{
\includegraphics[angle=-90,width=0.47\textwidth,clip]{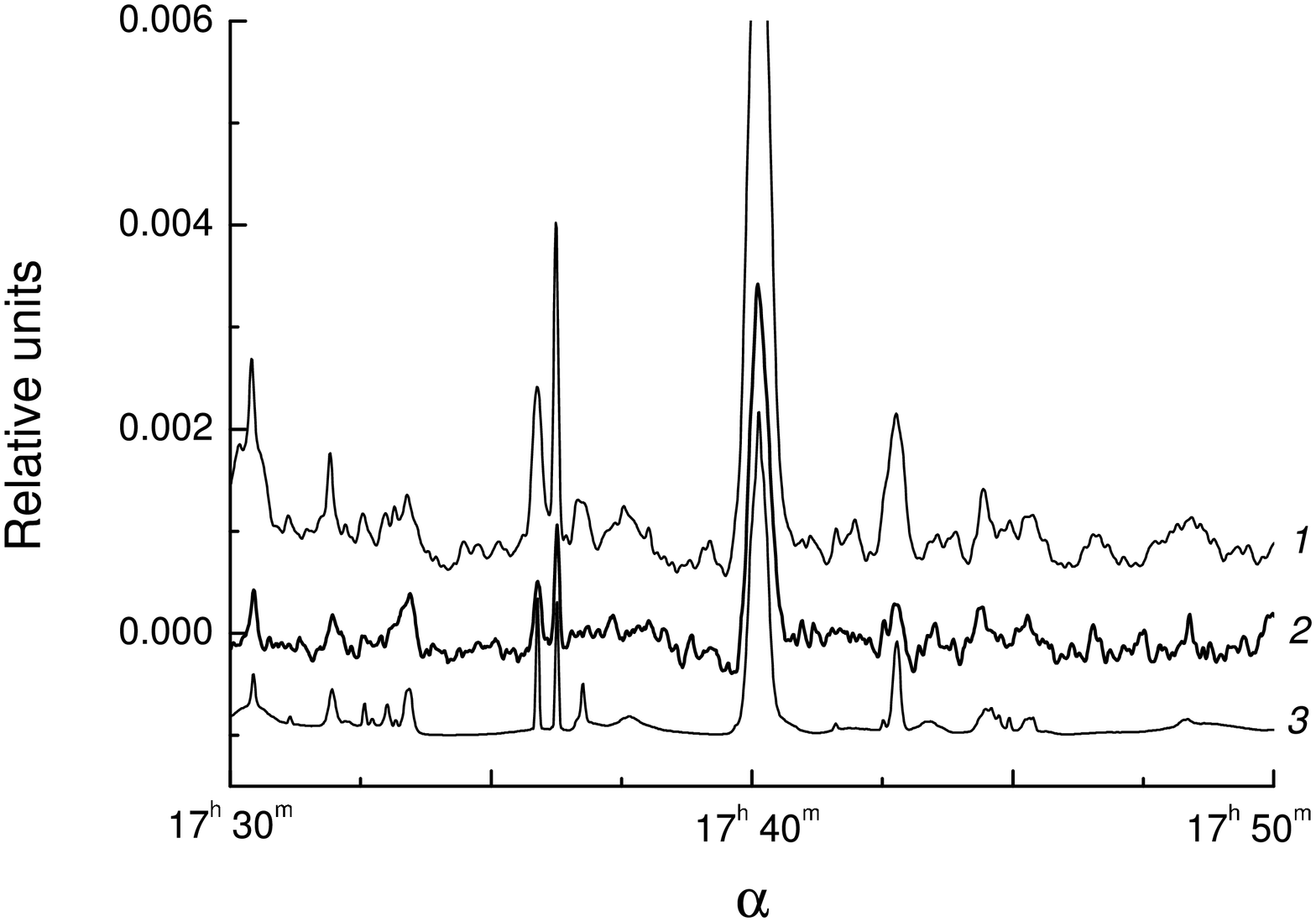}
}
}
}
\caption{
Normalized 20-minute simulated sky scans obtained using $4\degr\times 4\degr$ NVSS images
(curves \emph{1}); the corresponding real records of the RZF survey (curves \emph{2}) at
$\lambda7.6$~ÓÍ, and simulated scans based on the data of NVSS catalog [\cite{b1:Majorova_n}]
(curves \emph{3}). All scans are normalized to the 3C84 signal level. The curves are shown
shifted with respect to each other along the vertical axis.
}
\label{fig5:Majorova_n}
\end{figure}

It is of interest to compare the results of simulations not only
with real records, but also with simulated scans based on the
NVSS catalog [\cite{b1:Majorova_n}] data.
Figure~\ref{fig5:Majorova_n} features 20-minute simulated sky
scans obtained using NVSS images (curves \emph{1}) and simulated scans
based on the data of the NVSS catalog (curves \emph{3}), as well
as the corresponding averaged real records of the RZF survey
(curves \emph{2}) at 7.6~cm. All scans are normalized to the
level of the 3C84 signal. The simulated scans based on the data
of NVSS catalog and averaged real records were kindly provided by
N.~N.~Bursov. Records are shifted with respect to each other
along the vertical axis.

It is evident from a comparison of the curves shown in
Fig.~\ref{fig5:Majorova_n} that both models agree well with
observational data, however, the small-scale structure of
simulated scans obtained using NVSS images is closer to that of
real sky records. The simulated scans based on the NVSS catalog
are smoother as compared to curves \emph{1} and \emph{2}.

\begin{figure*}[tbp]
\centerline{
\vbox{
\hbox{
\includegraphics[angle=-90,width=0.40\textwidth,clip]{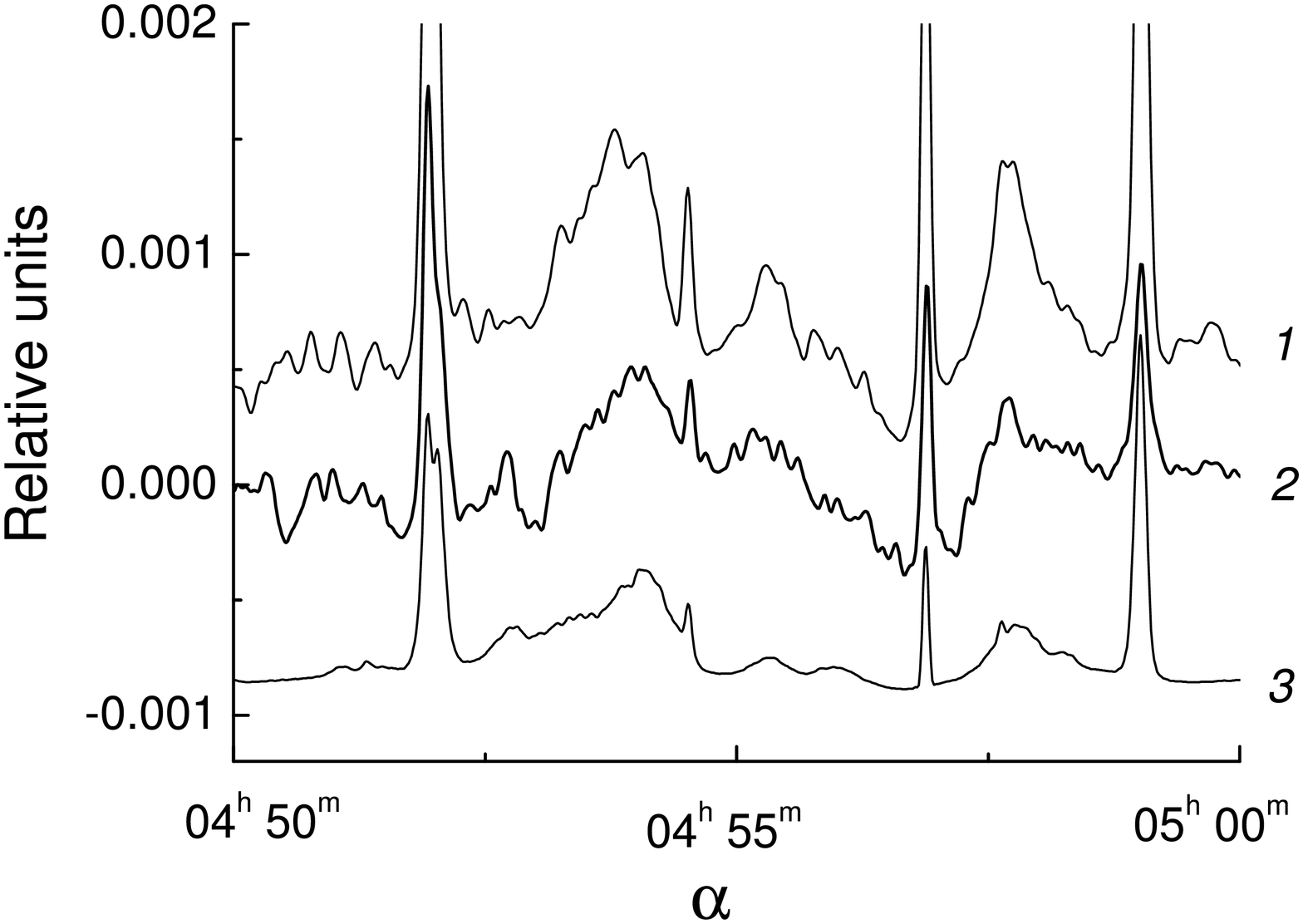}
\includegraphics[angle=-90,width=0.40\textwidth,clip]{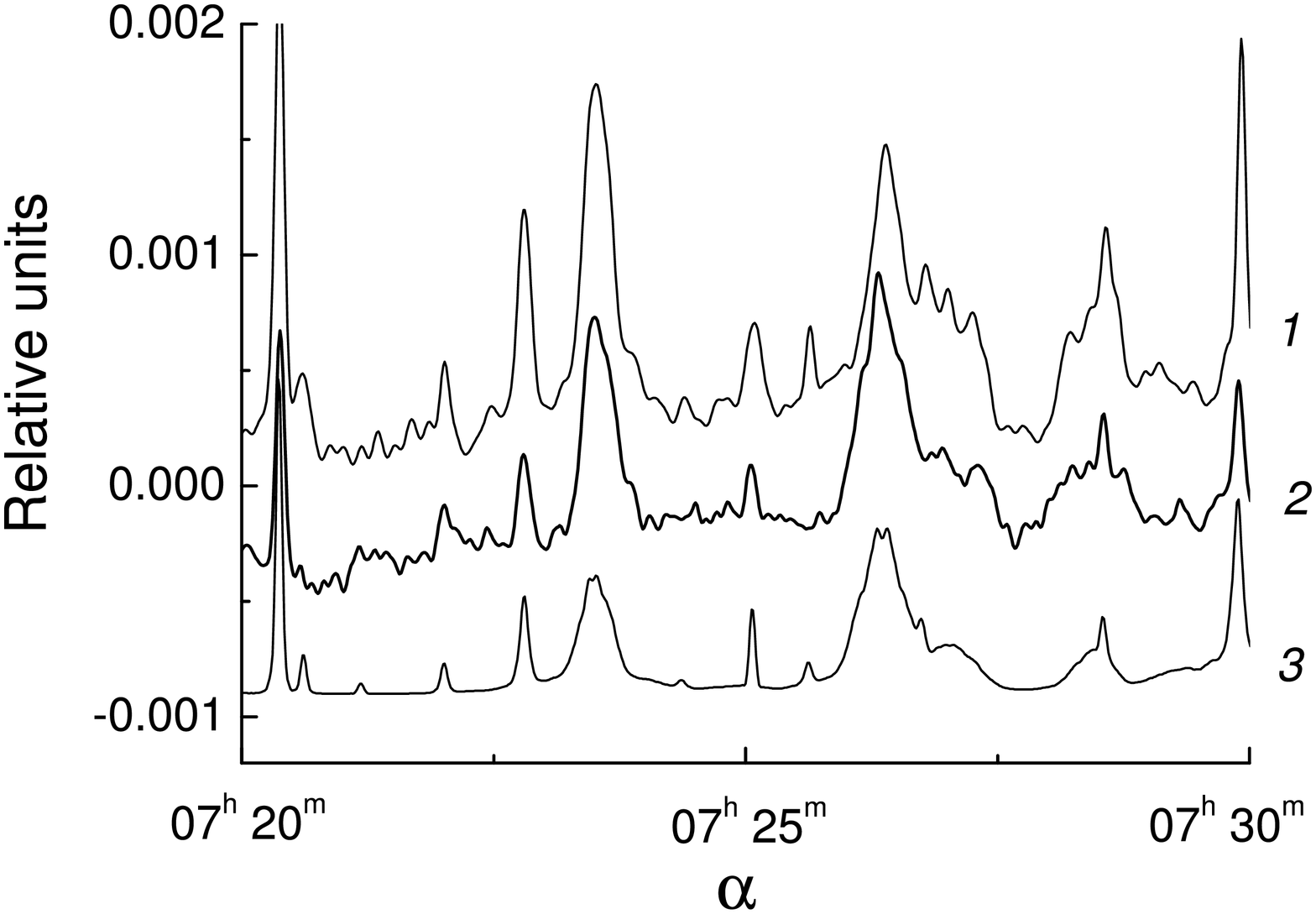}
}
\hbox{
\includegraphics[angle=-90,width=0.40\textwidth,clip]{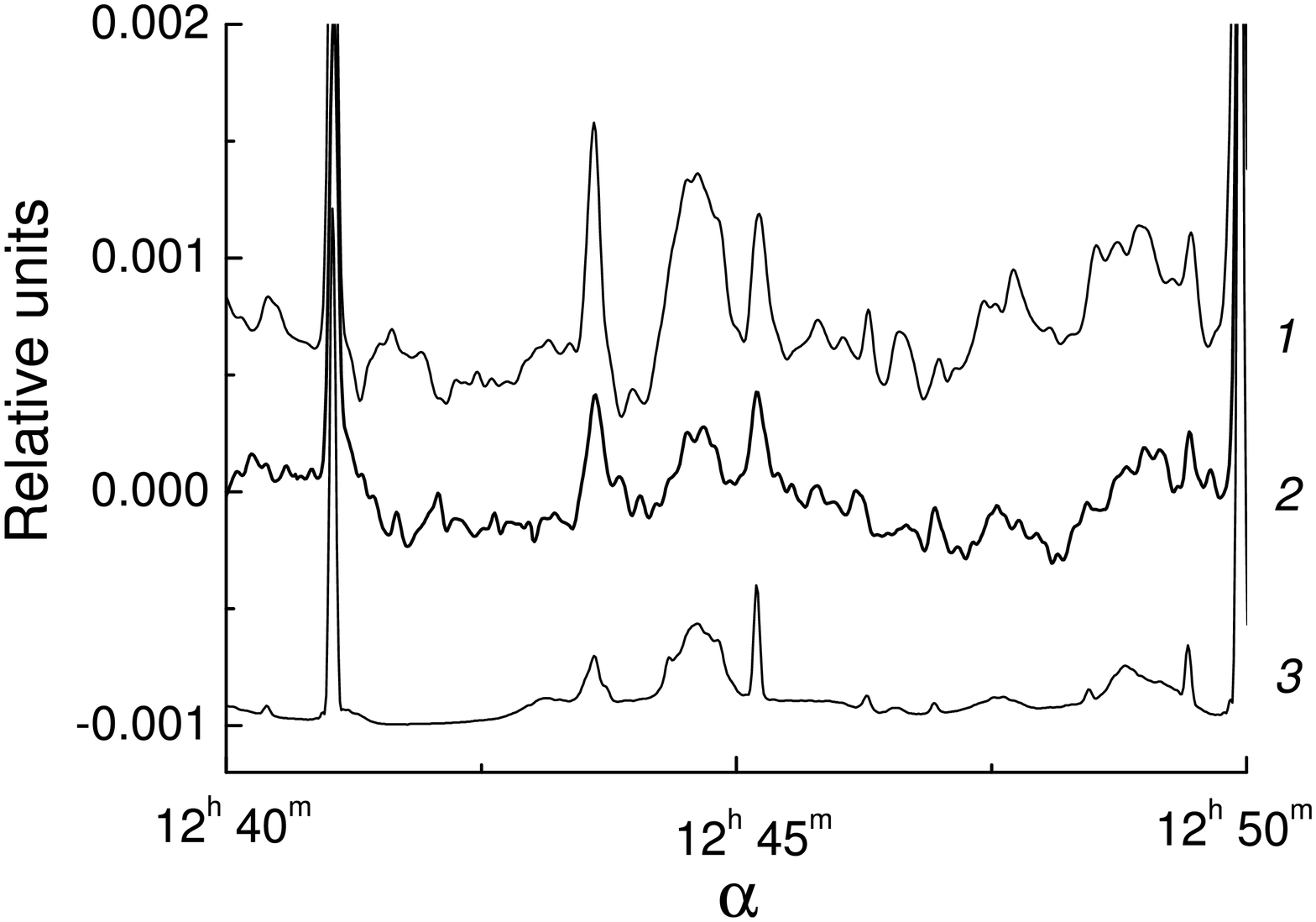}
\includegraphics[angle=-90,width=0.40\textwidth,clip]{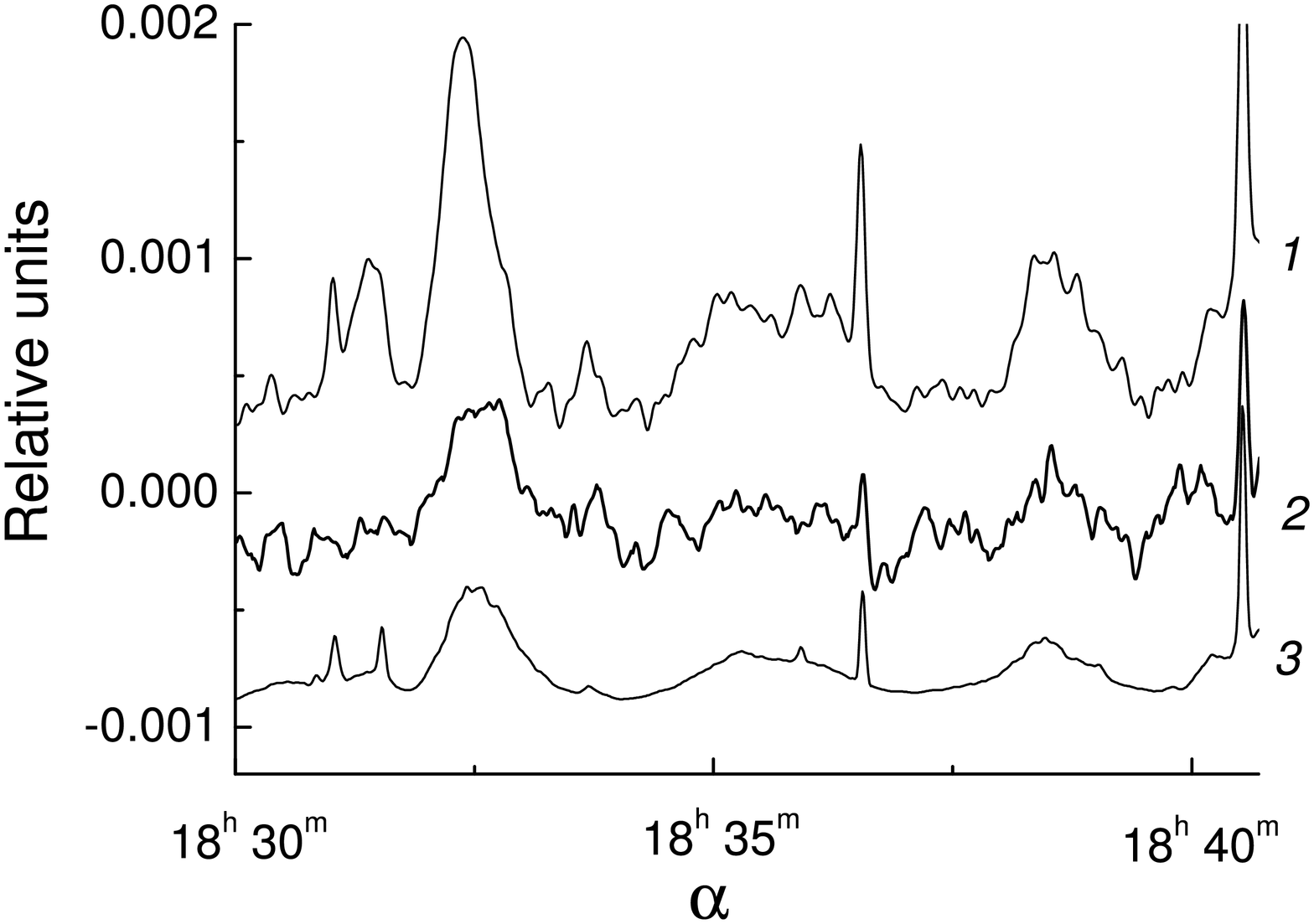}
} } }
\caption{
Ten-minute portions of normalized simulated sky scans obtained using
$4\degr\times 4\degr$ NVSS images (curves \emph{1}); the
corresponding averaged records of the RZF survey  (curves
\emph{2}) at 7.6~cm, and simulated scans based on the data from
the NVSS catalog [\cite{b1:Majorova_n}]. }
\label{fig6:Majorova_n}
\end{figure*}

These differences are even more apparent in
Fig.~\ref{fig6:Majorova_n}, which shows 10-minute portions of
records. The differences can be explained by the fact that
simulated scans based on the data of the NVSS catalog contain
only  the NVSS sources in the given declination band that have
fluxes above a certain level. At the same time, the simulated
scans constructed using NVSS images contain, in addition to NVSS
sources, VLA detector noise convolved with the PB of RATAN-600, and
the sources whose signals are at the detection level and close to
the noise of the NVSS survey. Such sources could have escaped the
NVSS catalog.

Note that the simulation of the RZF survey using the technique
proposed by Bursov and Majorova [\cite{b1:Majorova_n}] also has
certain advantages. Thus the levels of the signals from the
sources with known spectral indices are closer to the actual
values for the model based on the NVSS catalog, because the
fluxes of these sources were converted from  $\lambda21$~cm to
$\lambda7.6$~cm in accordance with their spectral indices.

Such an individual  correction of source fluxes is very problematic to perform in the
model proposed in this paper. The fluxes can be converted only in bulk for all sources in the
record using the average spectral index.

\section{SOME PROBLEMS THAT CAN BE SOLVED BY SIMULATING THE RZF SURVEY}

Consider now the problems that can be solved with the simulated
scans of the RFZ sky survey obtained.

Note that simulated scans based both on the data of the NVSS
catalog and NVSS images have already been used  to identify
sources of the zenith-field sky survey and in the compilation of
the RZF catalog [\cite{p1:Majorova_n}]. The high degree of
correlation between simulated and observed scans ($\sim 90\%$)
allowed us to separate the contribution of discrete sources and
background radiation, and to identify the sources in the survey
records with the sources of the NVSS catalog. Simulated curves
are especially valuable for detecting weak objects and for finding
new objects undetected at decimeter-wave frequencies.

In this paper we also address a number of other problems. First,
the analysis of simulated scans based on NVSS images allows us to
estimate the accuracy of identification of sources and of the
determination of their fluxes on sky-transit records. Second,
simulated scans can be used to clean a real record from sources
and estimate the minimum residual noise. Furthermore, it is of
interest to estimate the spectral indices of the sources from the
ratio of their signals on real and simulated scans and to compare
their distribution with the distribution of spectral indices
computed by Bursov et al.~[\cite{p1:Majorova_n}].

Before we start addressing these problems, let us list the main
factors that may affect the accuracy of the simulation based on
NVSS images. These factors include: the accuracy to which we know
the power beam pattern  of the radio telescope; the correctness of
the source extraction on the simulated scans obtained, and
the effect of the sizes of NVSS image areas on the results of
simulation.

The values of the vertical PB are used to convert the antenna
temperatures of the sources to the corresponding antenna
temperatures for the central section of the survey and for the
subsequent determination of the source fluxes. Majorova and
Trushkin [\cite{m2:Majorova_n}] and Majorova and
Bursov [\cite{m3:Majorova_n}] show that if the antenna is in good
state, then the computed PB agrees with the experimental PB very
accurately and therefore the errors due to the use of the
computed PB are minimal and amount to mere $1\div 3\%$.

Extraction of sources on simulated and real scans using the
technique of Gauss analysis involves certain difficult aspects,
especially when it comes to weak sources and sources with close
right ascensions. We estimate the accuracy of the determination
of the fluxes of the sources identified on records in Paragraph~6.

As for the effect of the size of the  NVSS images on the
results of simulation, our main task is to determine the maximum
size of the area to minimize the simulation errors.

\begin{figure*}[t]
\centerline{
\hbox{
\includegraphics[angle=-90,width=0.33\textwidth,clip]{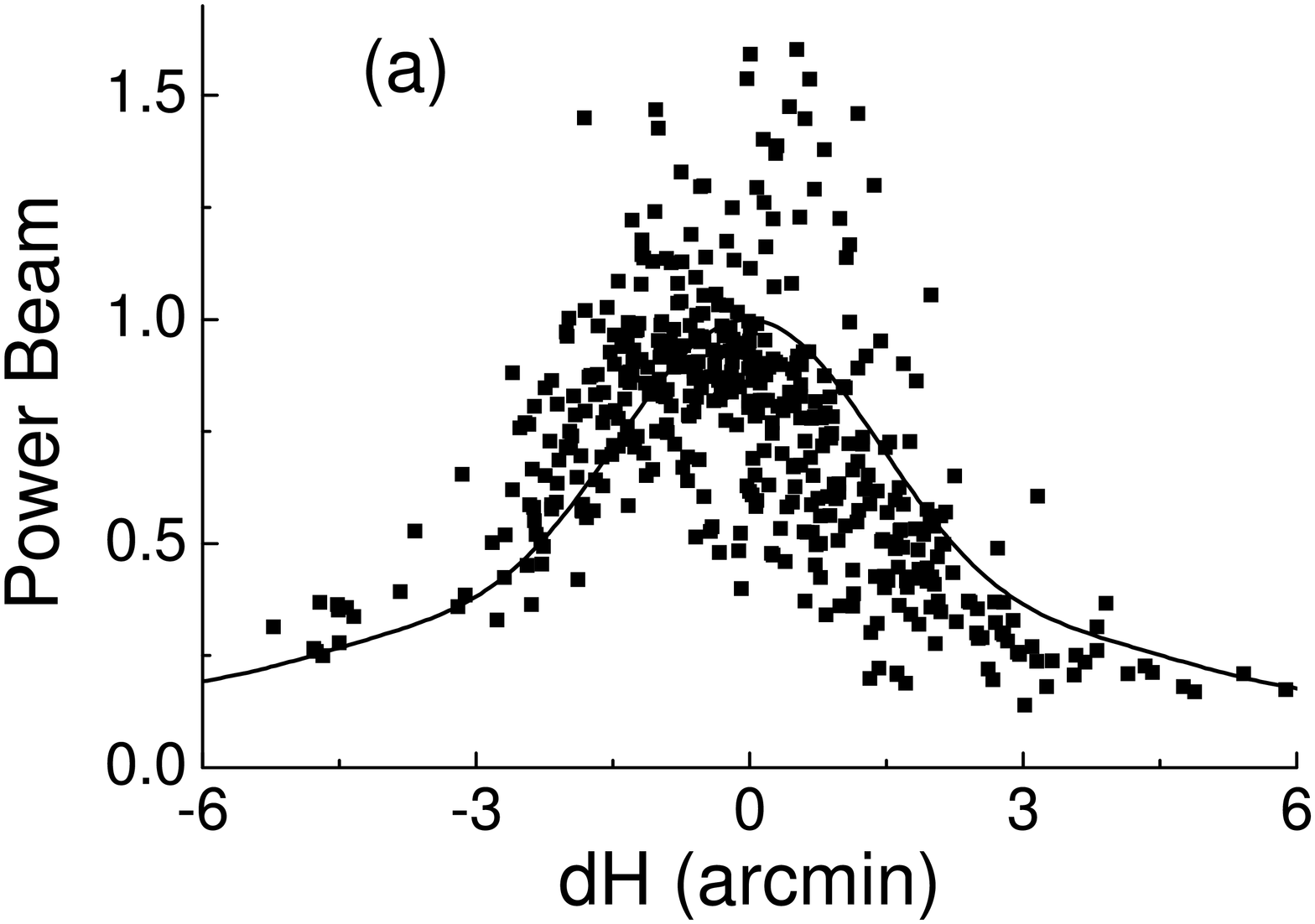}
\includegraphics[angle=-90,width=0.33\textwidth,clip]{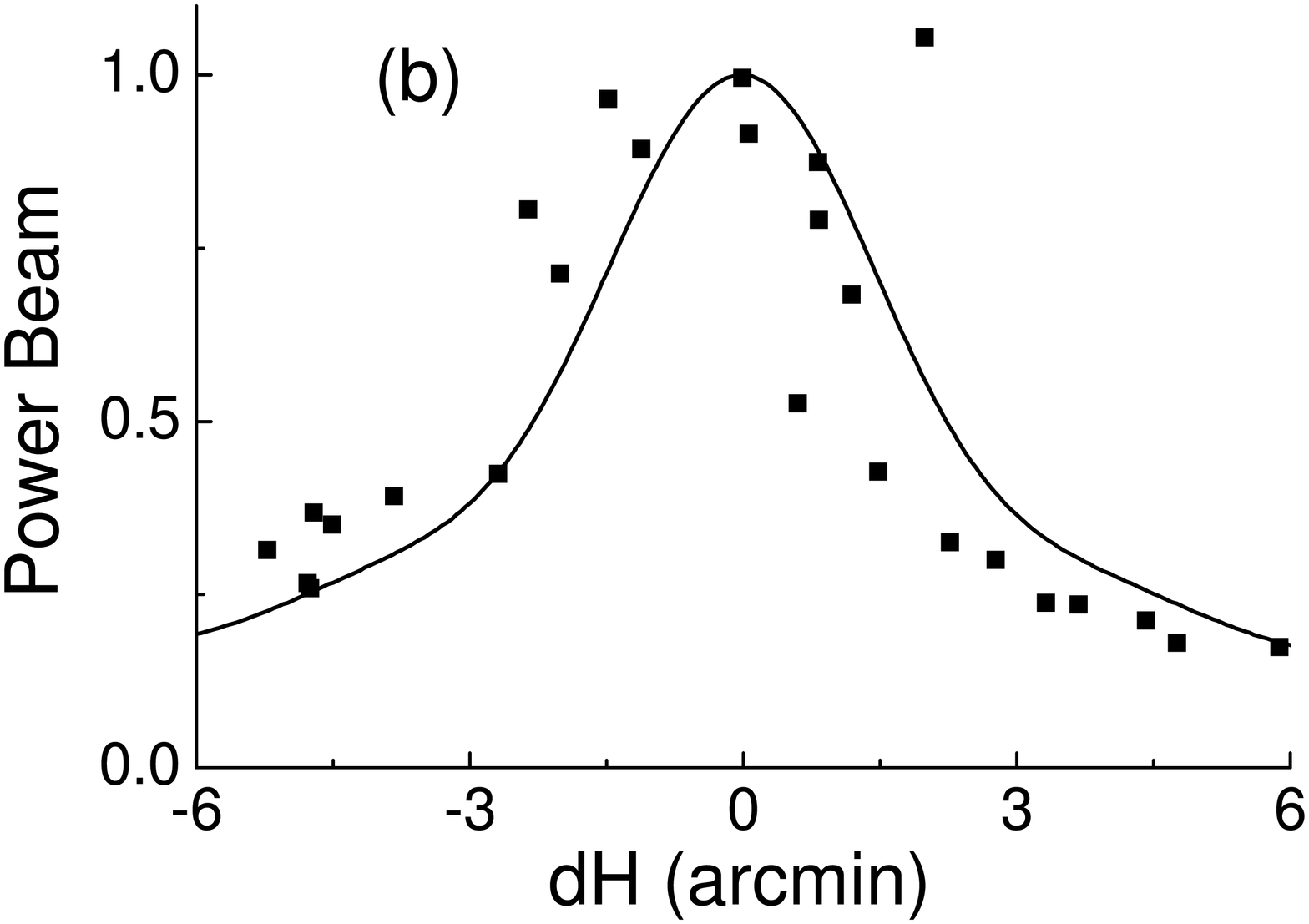}
}
}
\caption{
Vertical PB constructed from sources extracted on simulated scans (filled squares)
for the entire sample of sources ($S \ge 2.7$~mJy) (a) and from sources with fluxes  $S \ge
200$~mJy---(b). Simulated scans were obtained using  $4\degr\times 4\degr$ NVSS images.
The solid lines indicate the vertical section of the computed PB at the wavelength of 7.6~cm
after its convolving in each horizontal section with a Gaussian of width $\theta_{0.5} = 45''$.
}
\label{fig7:Majorova_n}
\end{figure*}

\begin{figure*}[tbp]
\centerline{
\hbox{
\includegraphics[angle=-90,width=0.33\textwidth,clip]{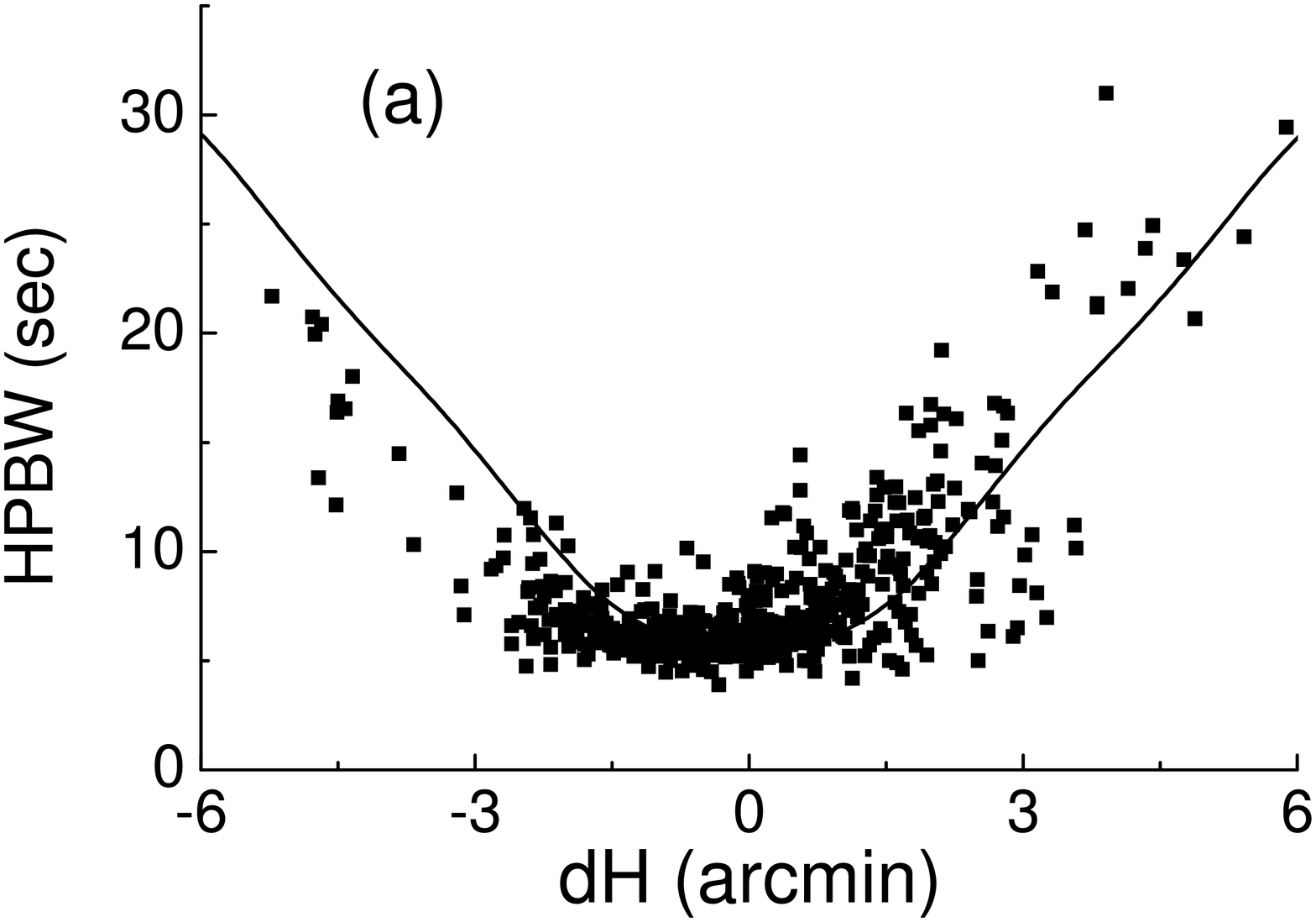}
\includegraphics[angle=-90,width=0.33\textwidth,clip]{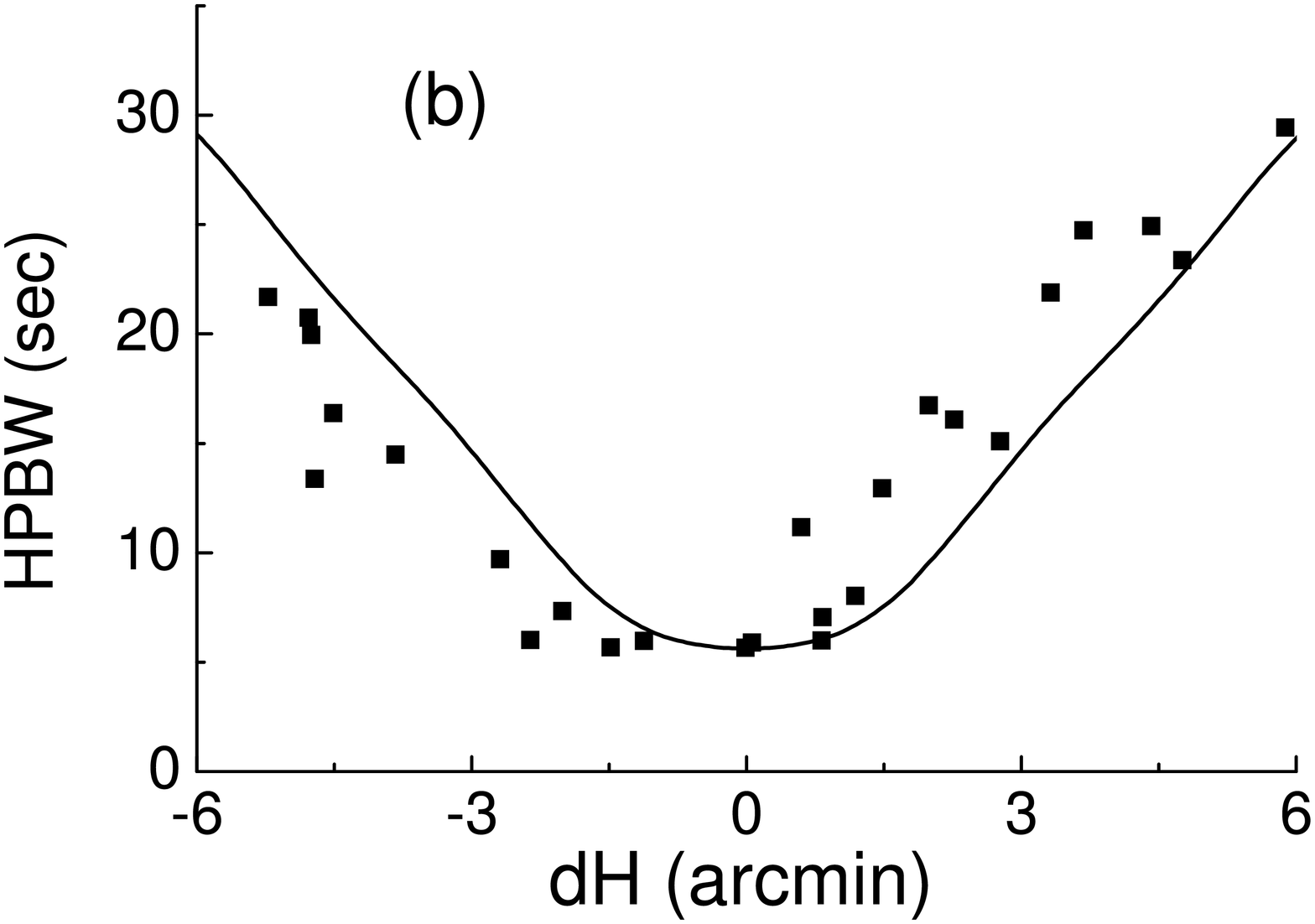}
} }
\caption{
The $dH$ dependences of the halfwidth  of PB constructed using the data
for the sources extracted on simulated scans (filled squares)
for the entire sample of sources ($S \ge 2.7$~mJy)---(a) and
using sources with fluxes $S \ge 200$~mJy---(b). Simulated scans
were obtained from  $4\degr\times 4\degr$ NVSS images. The solid lines
show the halfwidths of the computed PB at the wavelength of
7.6~cm convolved  in each horizontal section with a Gaussian with a
width of $\theta_{0.5} = 45''$. } \label{fig8:Majorova_n}
\end{figure*}

\begin{figure*}[tbp]
\centerline{
\hbox{
\includegraphics[angle=-90,width=0.30\textwidth,clip]{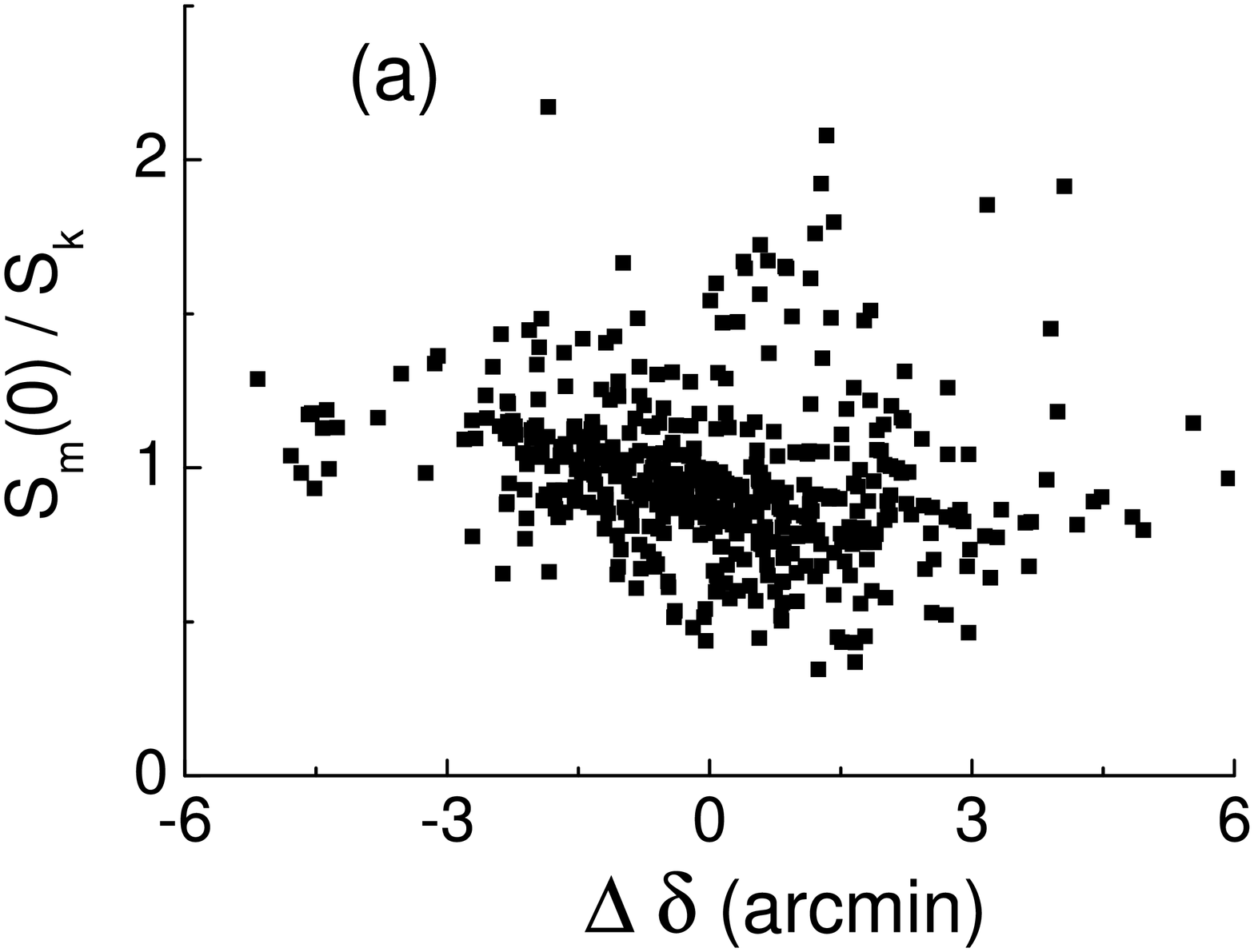}
\includegraphics[angle=-90,width=0.30\textwidth,clip]{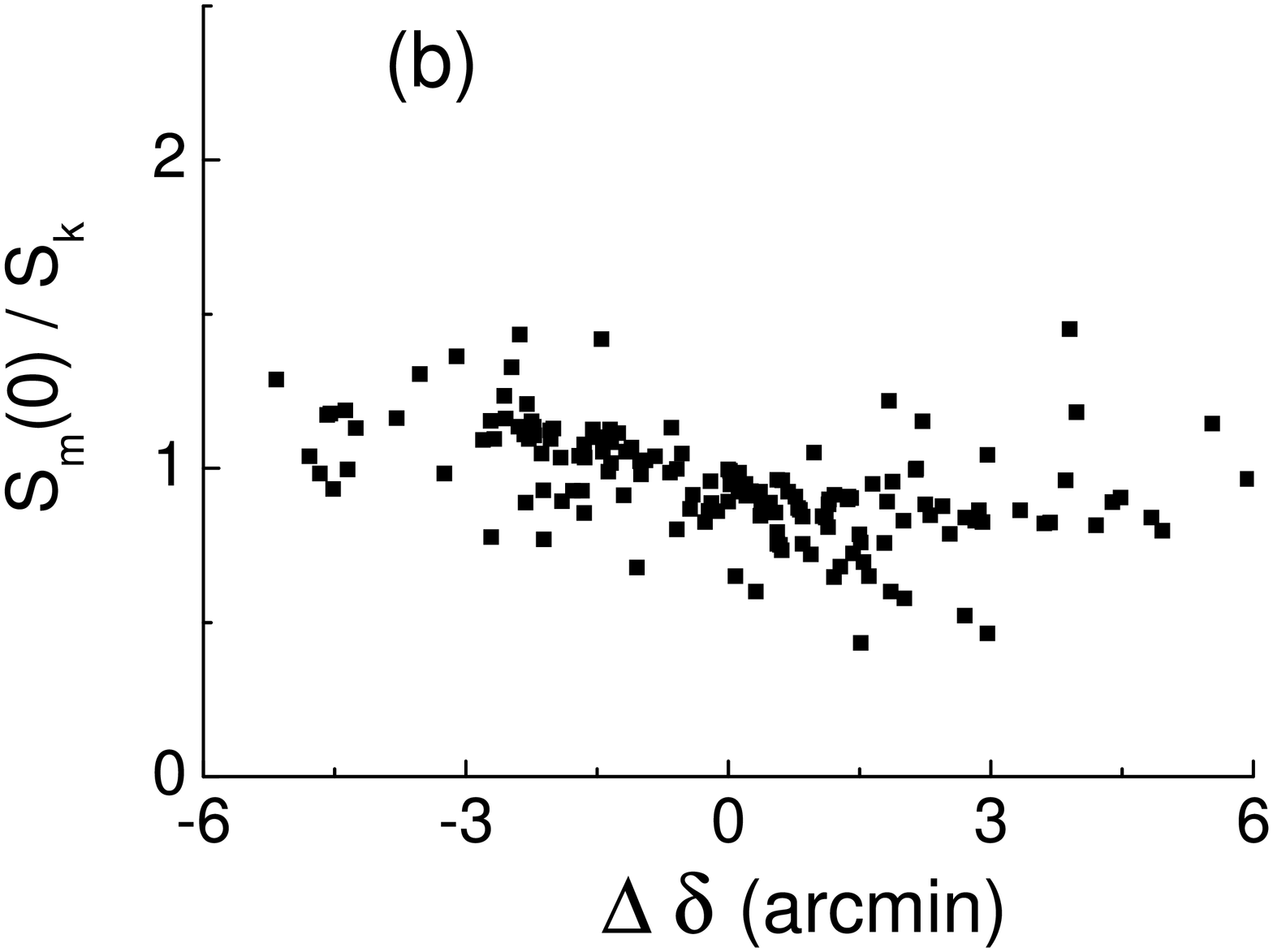}
\includegraphics[angle=-90,width=0.30\textwidth,clip]{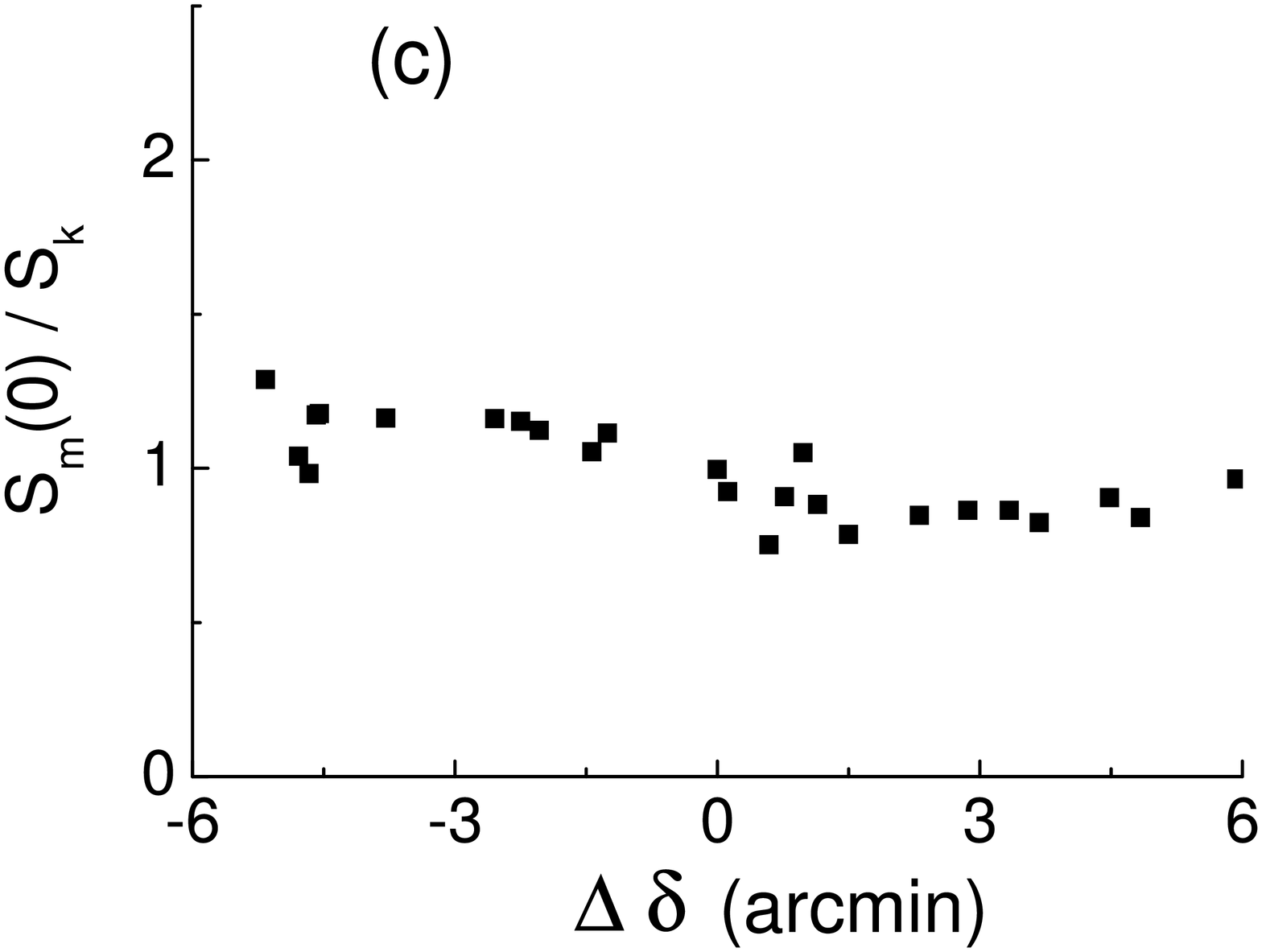}
}
}
\caption{
The $\Delta \delta$ dependences of $S_{m}(0)/S_{k}$ constructed using the entire
sample of sources extracted on simulated scans ($S \ge 2.7$~mJy) (a) and
using sources with fluxes $S \ge 30$~mJy (b) and $S \ge 200$~mJy (c). The simulated scans were
obtained from $4\degr\times 4\degr$ NVSS images.
}
\label{fig9:Majorova_n}
\end{figure*}

\begin{figure*}[tbp]
\centerline{
\hbox{
\includegraphics[angle=-90,width=0.30\textwidth,clip]{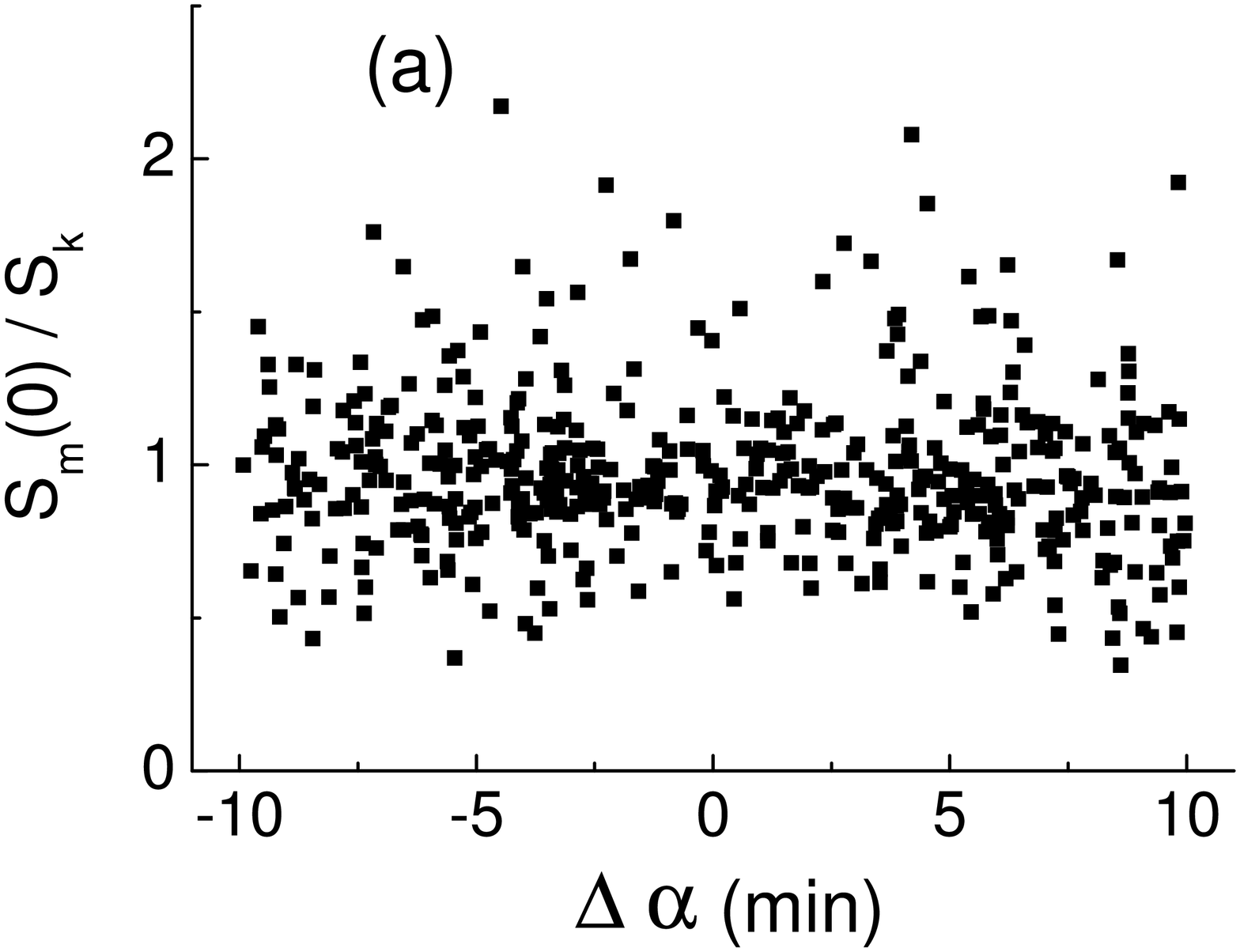}
\includegraphics[angle=-90,width=0.30\textwidth,clip]{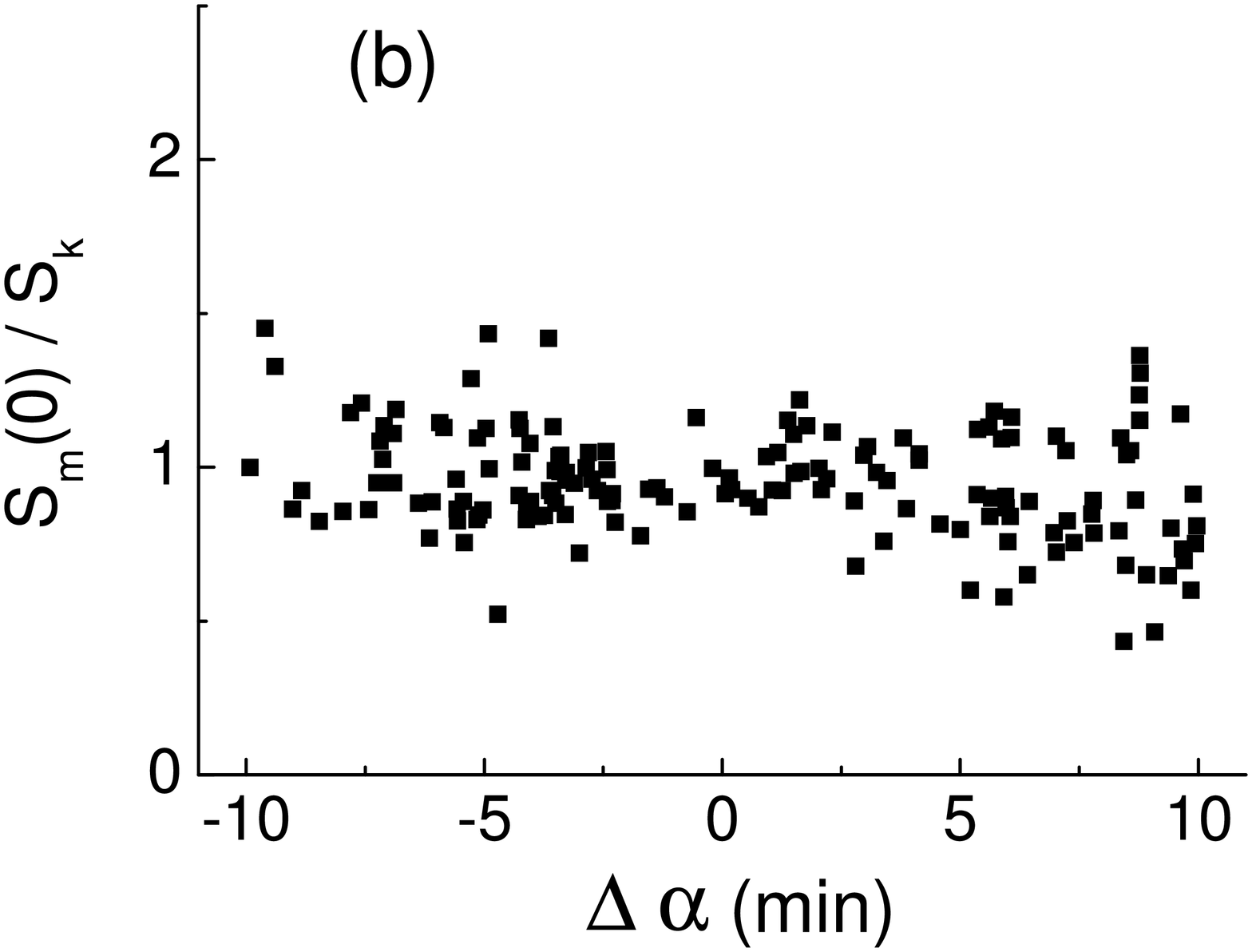}
\includegraphics[angle=-90,width=0.30\textwidth,clip]{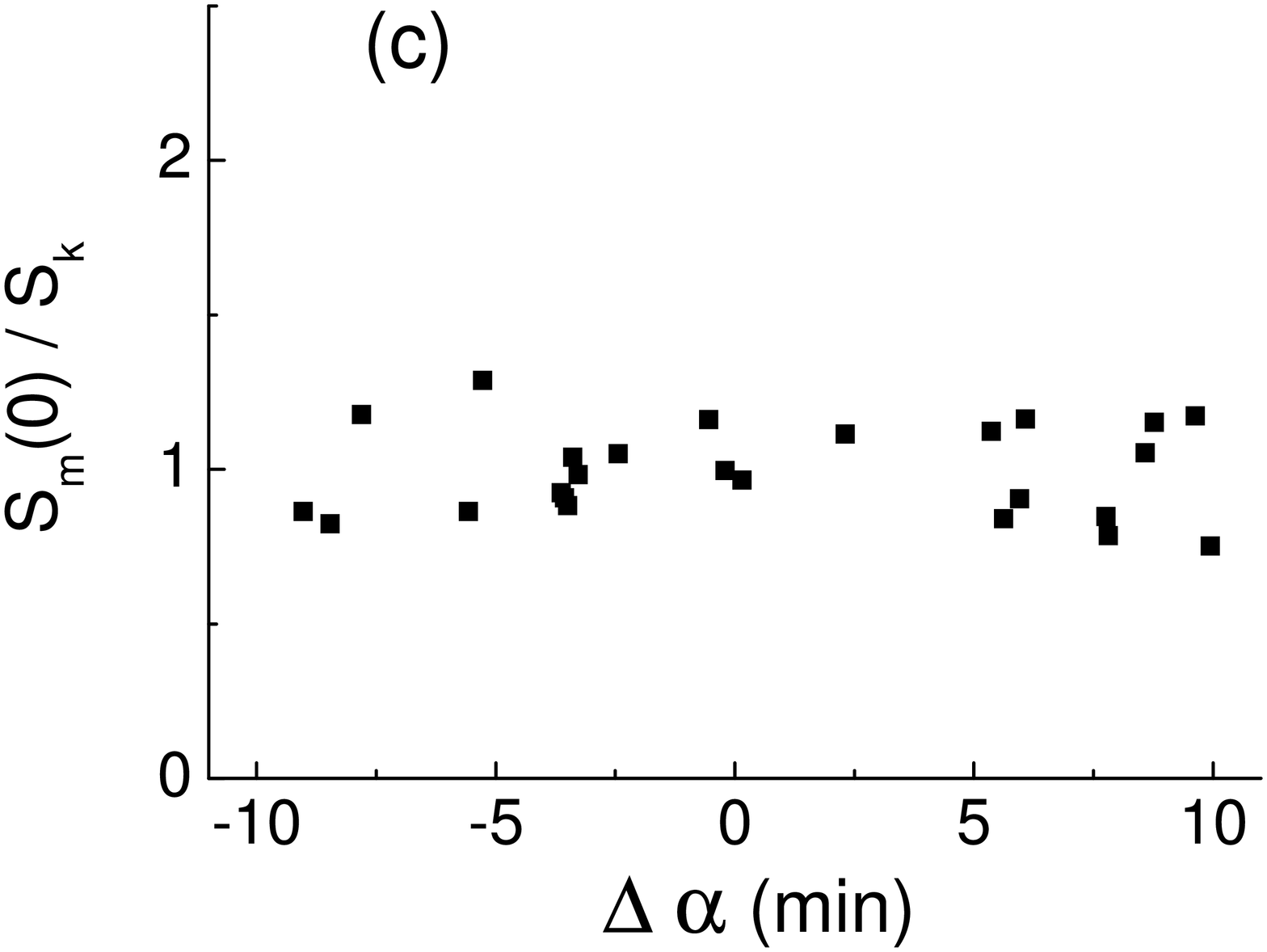}
}
}
\caption{
The $\Delta \alpha$ dependences of $S_{m}(0)/S_{k}$ constructed using the entire
sample of sources identified on simulated scans ($S \ge 2.7$~mJy) (a) and
using sources with fluxes $S \ge 30$~mJy (b) and $S \ge 200$~mJy (c). The simulated scans were
obtained from $4\degr\times 4\degr$ NVSS images.
}
\label{fig10:Majorova_n}
\end{figure*}

\section{EFFECT OF THE SIZE OF NVSS IMAGES ON  THE SIMULATION RESULTS}

To perform the task mentioned above using simulated scans based on
7.6-cm NVSS images, we identify the sources of the NVSS catalog
located within the $\pm 6'$-band of the central declination of
the survey. Like in the case of real records, we use the Gauss
analysis [\cite{p1:Majorova_n,b2:Majorova_n}] to identify the
sources, and control the declinations of the sources by the
halfwidths of the Gaussians. To this end, we use the computed
dependences of the halfwidth of the PB on its distance from the
central section.

We identified a total of about 500 NVSS catalog sources with 21-cm fluxes $S > 2.7$~mJy.
We determined the signal level for each of the sources identified both in relative units
(in terms of the 3C84 signal level) and absolute units (mJy).

\begin{figure*}[]
\centerline{
\vbox{
\hbox{
\includegraphics[angle=-90,width=0.40\textwidth,clip]{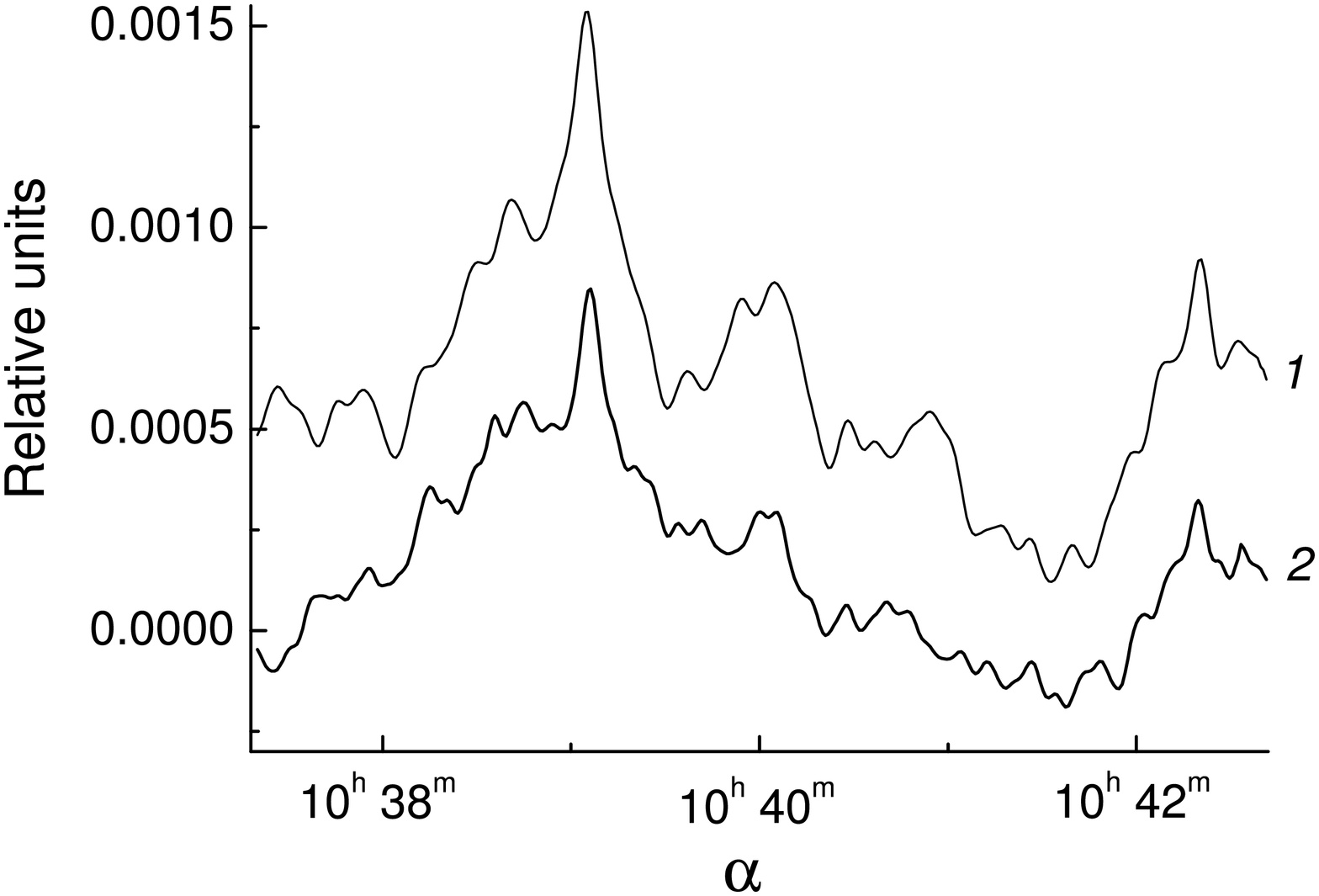}
\includegraphics[angle=-90,width=0.40\textwidth,clip]{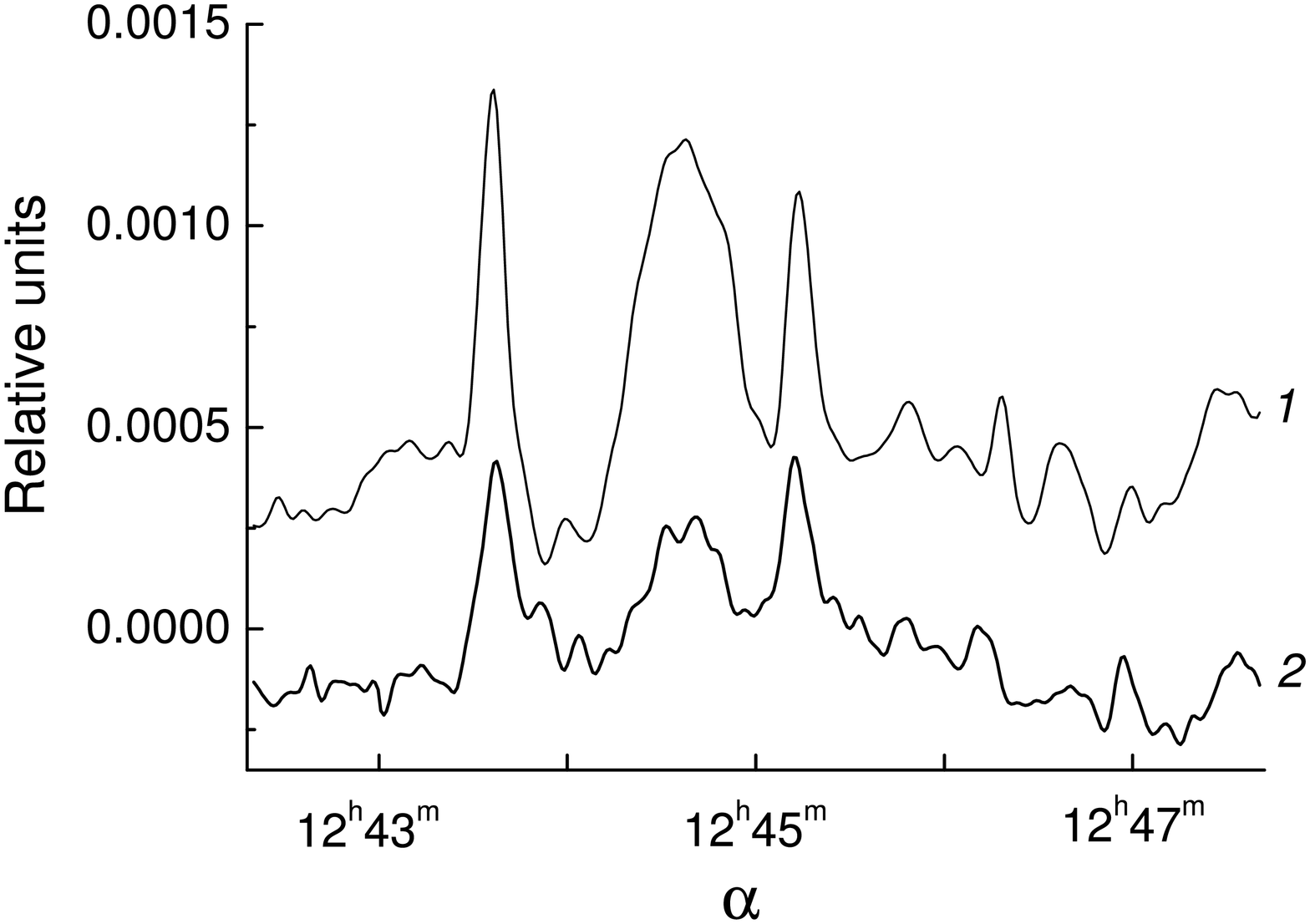}
}
\hbox{
\includegraphics[angle=-90,width=0.40\textwidth,clip]{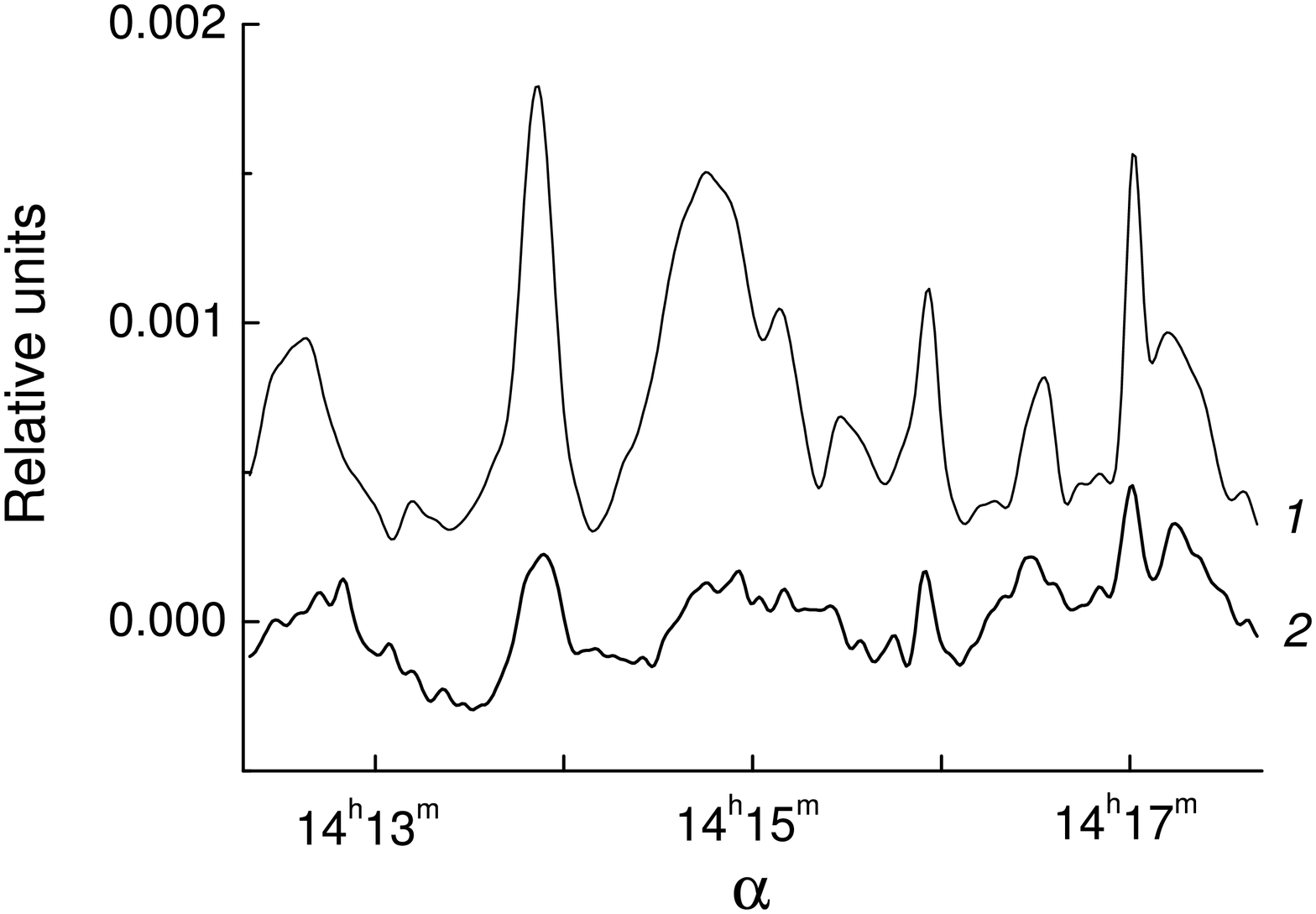}
\includegraphics[angle=-90,width=0.40\textwidth,clip]{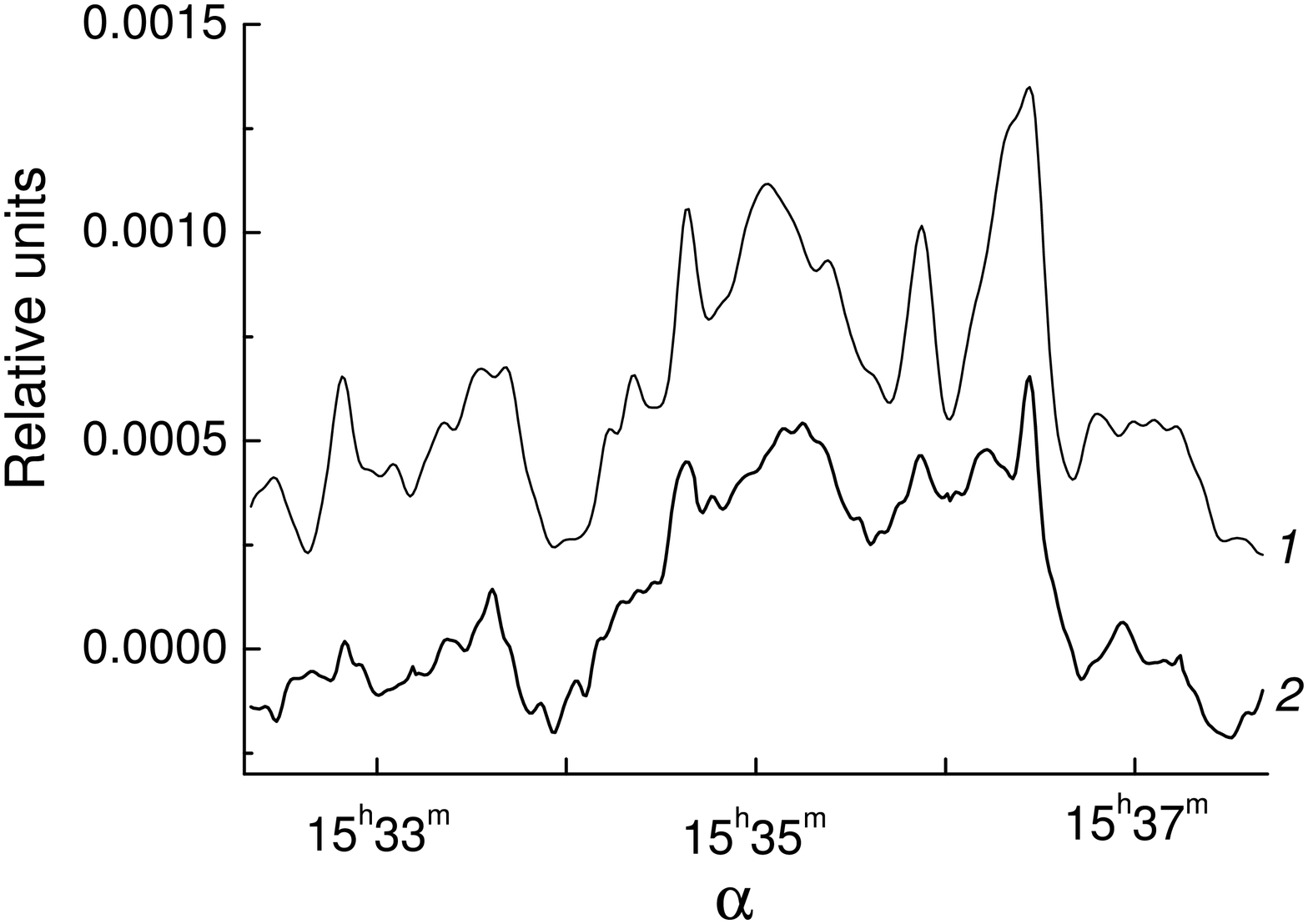}
} } }
\caption{
Normalized five-minute simulated sky scans obtained using  $1\degr\times 1\degr$
NVSS images (curves \emph{1}) and the corresponding averaged real
records of the RFZ survey (curves \emph{2}) at the wavelength of
7.6~cm. All scans are normalized to the 3ó84 signal level. The
curves are shown shifted along the vertical axis. }
\label{fig11:Majorova_n}
\end{figure*}

We construct the $\Delta \delta$ and $\Delta \alpha$ dependences
of $S_{m}(0)/S_{k}$ for this sample of sources and  determine the
vertical PB of RATAN-600 and the $dH$ dependence of the halfwidth
of the power beam. Here $S_{k}$ is the flux of the
source according to the NVSS catalog; $S_{m}(0)$, the flux of the
same source as inferred from the simulated scan and reduced to
the central section of the survey ($\Delta \delta = 0$), and $dH$
is the elevation offset with respect to the central section of
the PB.

\begin{equation}
S_{m}(0) = S_{m}(\Delta \delta)/F_{r}(dH),
\end{equation}

\noindent where $S_{m}(\Delta \delta)$ is the flux of the source
with the coordinates ($\delta_{ist}$, $\alpha_{ist}$) as inferred
from the simulated scan; $F_{r}(dH)$, the vertical PB computed
using computer codes in accordance with the technique described by
Majorova [\cite{m1:Majorova_n}], $\Delta \delta = \delta_{ist} -
\delta_{0}$, $\Delta\alpha = \alpha_{ist} - \alpha_{0}$.

During the survey considered the central section of the PB passes through the central section
of the survey, implying that $dH = \Delta \delta$.

The values $F_{Å}$ of the vertical PB of the radio telescope for various  $dH$ can be inferred
from the following formula:

\begin{equation}
 F_{Å}(dH) = S_{m}(\Delta \delta)/S_{k},
\end{equation}

\noindent where $dH = \Delta \delta$, using the source fluxes
inferred from simulated scans and the fluxes of the same sources
listed in the NVSS catalog.

The halfwidths of the power  beams are equal to the halfwidths of the Gaussians
fitted when identifying sources on simulated scans.

The filled squares in Figs.~\ref{fig7:Majorova_n}(a) and
\ref{fig8:Majorova_n}(a) show the vertical power beams
and the $dH$ dependences of HPBW constructed from the entire sample
of sources identified on simulated scans, and
Figs.~\ref{fig7:Majorova_n}(b) and \ref{fig8:Majorova_n}(b) show
the same power beams and dependences, but constructed from
bright sources with fluxes $S \ge 200$~mJy. The solid lines show
the vertical PB and the computed $dH$ dependence of HPBW. Note
that the two-dimensional power beams were convolved  with
the images already convolved with the instrumental function of
the VLA radio telescope, and therefore we convolved  the computed PB
of RATAN-600 in each of the horizontal sections with a Gaussian of
halfwidth equal to $\theta_{0.5}=45''$ (HPBW VLA
[\cite{co1:Majorova_n}]).

A sufficiently strong discrepancy between the power beams
 constructed from  simulated scans and the computed curves is
immediately apparent from Figs.~\ref{fig7:Majorova_n} and
\ref{fig8:Majorova_n}. The pattern of the discrepancy does not
depend on the brightness of the sources used to construct the
dependences in question.

Figure~\ref{fig9:Majorova_n} shows the $\Delta \delta$
dependences of $S_{m}(0)/S_{k}$ constructed ~~for~~ the~~ entire~~
sample~~ of~ sources  \mbox{($S \ge 2.7$~mJy)} (a) and for the subsamples of
sources with fluxes $S \ge 30$~mJy (b) and  $S \ge 200$~mJy (c).
All these dependences have a well-defined slope, indicating that
the source fluxes inferred from simulated scans differ from the
fluxes of the same sources listed in the NVSS catalog, and the
discrepancy increases with the difference between the
declinations of the source and of the central section of the
survey.  And, finally, the $\Delta\alpha$ dependences of
$S_{m}(0)/S_{k}$ shown in Fig.~\ref{fig10:Majorova_n} demonstrate
that the discrepancy between $S_{m}(0)$ and $S_{k}$ increases
with increasing distance between the source and the scan center
(or between the source and the centers of image areas Ihh20P40,
Ihh00P40, and Ihh40P40). These features are especially apparent
for the dependences based on the data for strong sources with
fluxes $S \ge 30$~mJy and \mbox{$S \ge 200$~mJy.}

All these discrepancies are due to projection of the sky points (i.e. points of a sphere)
onto the planes of the NVSS  images. Condon et
al. [\cite{co1:Majorova_n}] give the formulas that relate celestial
($\alpha, \delta$) coordinates to pixel coordinates ($x, y$) of
the image area:
\begin{equation}
x = x_{0} - cos\delta sin(\alpha - \alpha_{0})/\epsilon , \\
\end{equation}
\begin{equation}
y = y_{0} + [sin\delta sin\delta_{0} - cos\delta sin\delta_{0}cos(\alpha - \alpha_{0}]/\epsilon ,
\label{:Majorova_n}
\end{equation}

\noindent where $\epsilon = \pi /43200$~radian = $15''$,
$x_{0}=512, y_{0}=513$.

It is evident from the above formulas that the farther is the source from the center of the
sky area ($\alpha_{0}, \delta_{0}$), the greater is the error of its inferred position in the
image plane. The difference between values of $\Delta \delta$ and $\Delta y = y_{ist} - y_{0}$
results in
systematic errors in the inferred source fluxes reduced to the central section of the survey
using the vertical power beam  of the telescope. These errors, in turn, result in
non-zero slopes of the dependences shown in Fig.~\ref{fig9:Majorova_n}, deviation of the
simulated data points from the computed curves in Figs.~\ref{fig7:Majorova_n} and
~\ref{fig8:Majorova_n}, and in the increase of the scatter of data points toward the scan
boundaries in Fig.~\ref{fig10:Majorova_n}.

The dependences shown in Fig.~\ref{fig10:Majorova_n} can be used
to select the maximum right-ascension sizes of image areas such
that the  $S_{m}(0)/S_{k} \sim 1$ condition is satisfied
throughout the entire length of the simulated scan. This size is
on the order of  $d\alpha = \pm 2.5^{m}$ (or \mbox{$\sim 1\degr$).} The
declination size of the images, which is used in simulations, is
close to its optimum value. Thus for the wavelength of
$\lambda7.6$~cm it is  $d\delta = \pm37'15''$.

Thus the simulated scans obtained by convolving 20-minute (in right
ascension) sky areas of the NVSS survey with the two-dimensional PB
of RATAN-600 are of great interest for the extraction of
sources on real records of the zenith-field survey. The simulated
and real scans are highly correlated and this correlation is
especially useful for identifying weak sources. However, they are
not suitable for quantitative estimates. To this end, the size
of the image areas must be on the order of $1\degr\times 1\degr$
(about $5^{m}$ in terms of right ascension). Note that with such
an image size both the number of images for the 7.6-cm wavelength
and the number of simulated scans increases by a factor of four.
This was one of the reasons why we began our simulations with
$4\degr\times 4\degr$ NVSS image areas.

\section{SIMULATION OF THE ZENITH-FIELD SURVEY USING $1\degr\times1\degr$
NVSS IMAGES}

We simulated the zenith-field survey using  \mbox{$1\degr\times 1\degr$}
NVSS images at the wavelength of 7.6~cm. We downloaded the images
of the sky areas in FITS format from the
site of SkyView  [\cite{na2:Majorova_n}] and converted them into the F format
and then into 300$\times$300 binary matrices. We then convolved  the
latter with the two-dimensional PB of RATAN-600 computed for
$\lambda 7.6$~cm and having the same mesh size in  $\alpha$ and
$\delta$ as the images of the NVSS survey.

The centers of the image areas had the following coordinates: $\alpha_{0} = H^{h}5^{m}n$,
$\delta_{0} = +41\degr30'45'' + \Delta$, where $H$ are hours from 0 to 23,
$n = 0,1,2, ...11$, and $\Delta$ is the precession correction.

As a result, we obtained a total of 288 five-minute simulated
scans. Figure.~\ref{fig11:Majorova_n} (curves \emph{1}) shows
examples of such scans. The same figure shows the averaged
five-minute records of sky transits obtained in the RFZ survey at
the wavelength of 7.6~cm during the 1998--1999 period (curves
\emph{2}). The simulated and real scans shown in the figures are
normalized to the 3ó84 signal level. The curves are shifted with
respect to each other along the vertical axis.

As is evident from the figures, the simulated and observed scans are highly correlated.
The correlation coefficient is  $\sim 0.95$, which is higher than the correlation
coefficient between 20-minute simulated scans and the scans of the RFZ survey, and than the
correlation coefficient between the simulated scans based on the data of the NVSS catalog and
real records.

\begin{figure*}[tbp]
\centerline{
\hbox{
\includegraphics[angle=-90,width=0.33\textwidth,clip]{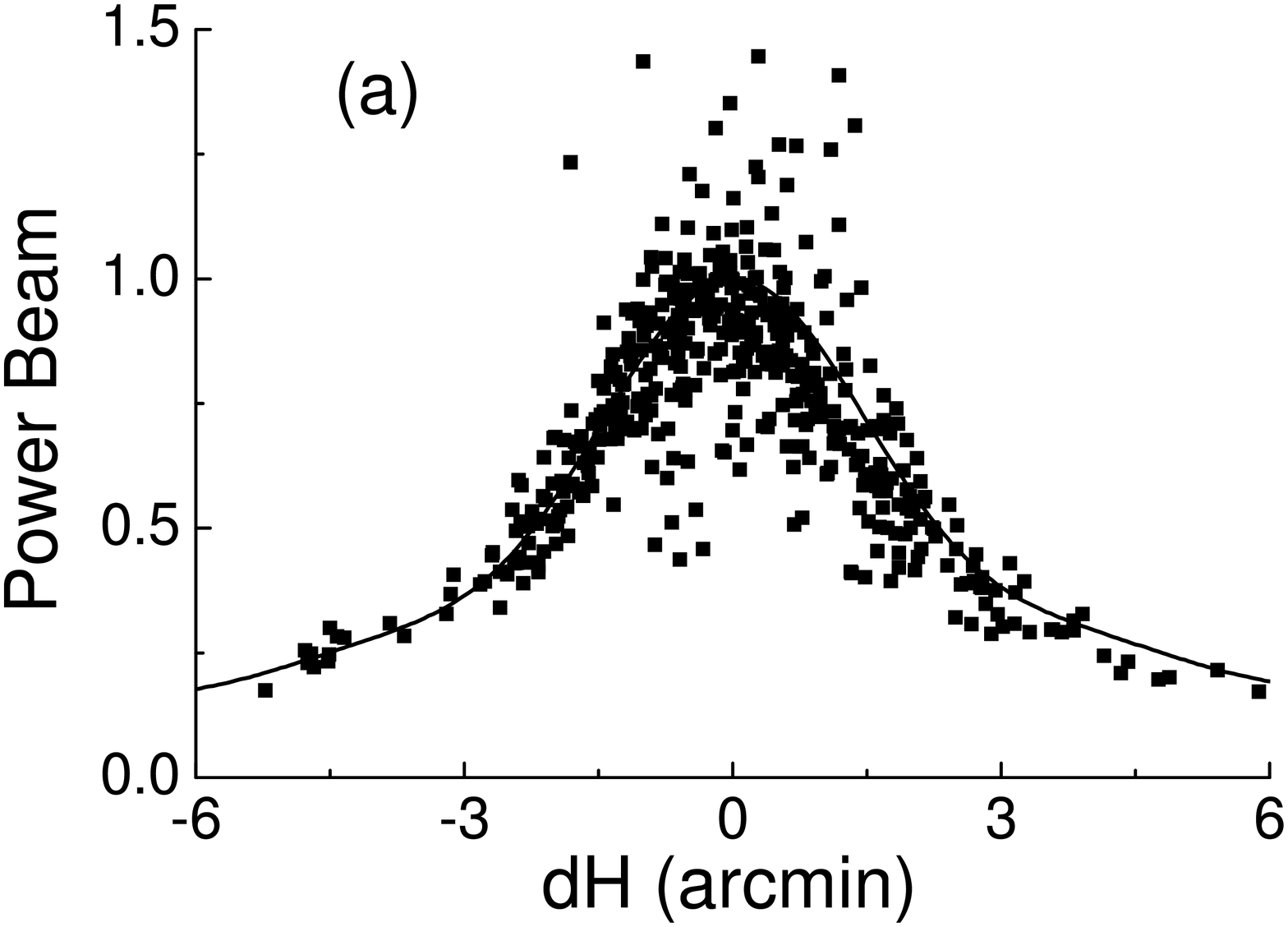}
\includegraphics[angle=-90,width=0.33\textwidth,clip]{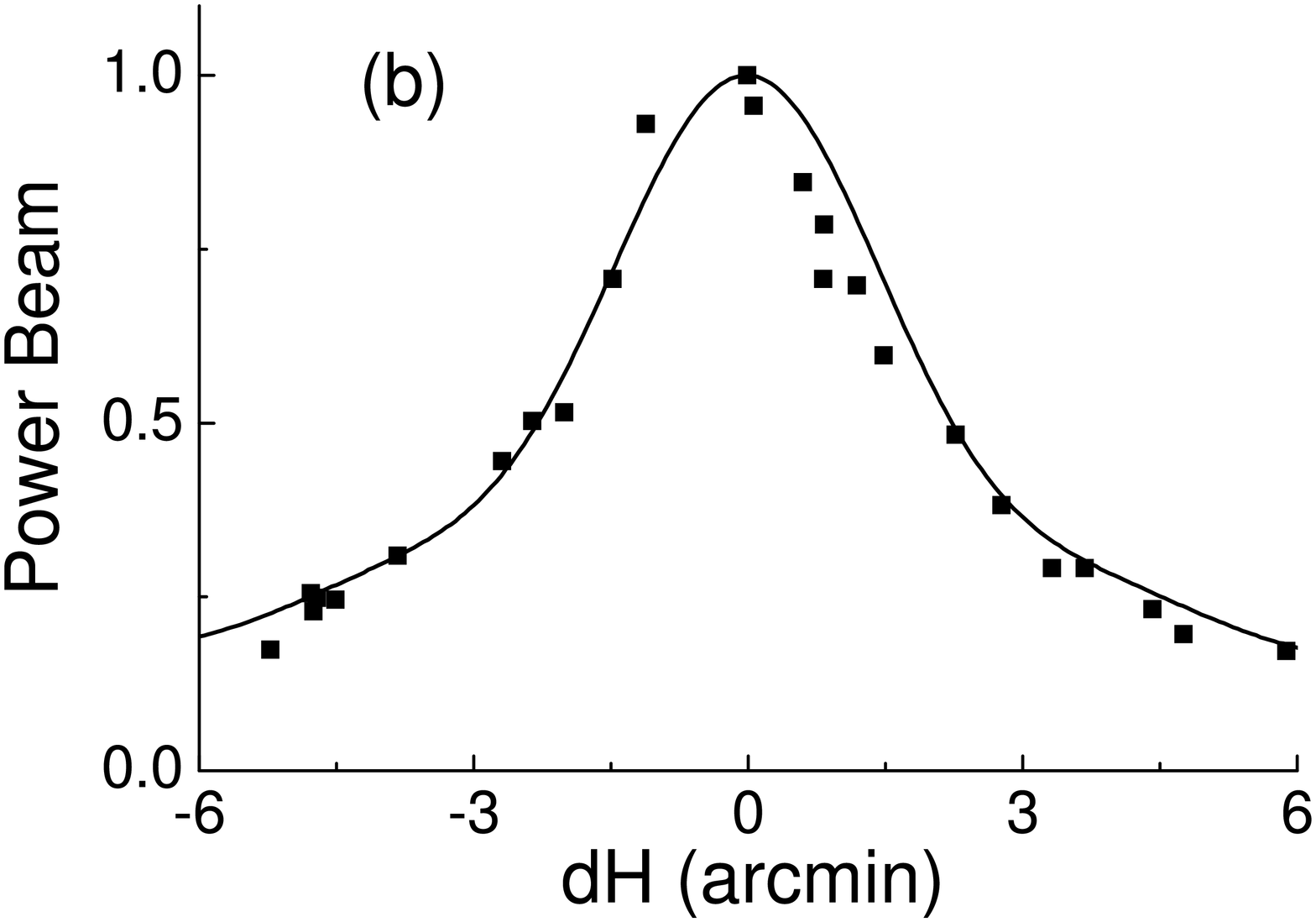}
}
}
\caption{
Vertical PB constructed using the sources extracted on simulated scans (filled squares):
for the entire sample of sources ($S \ge 2.7$~mJy) (a) and for sources with fluxes
$S \ge 200$~mJy (b). The simulated scans were obtained from   $1\degr\times 1\degr$ NVSS
images. The solid lines show the vertical section of the computed PB at 7.6~cm after its
convolving in each horizontal section with a Gaussian of size $\theta_{0.5} = 45^{''}$.
}
\label{fig12:Majorova_n}
\end{figure*}

\begin{figure*}[tbp]
\centerline{
\hbox{
\includegraphics[angle=-90,width=0.33\textwidth,clip]{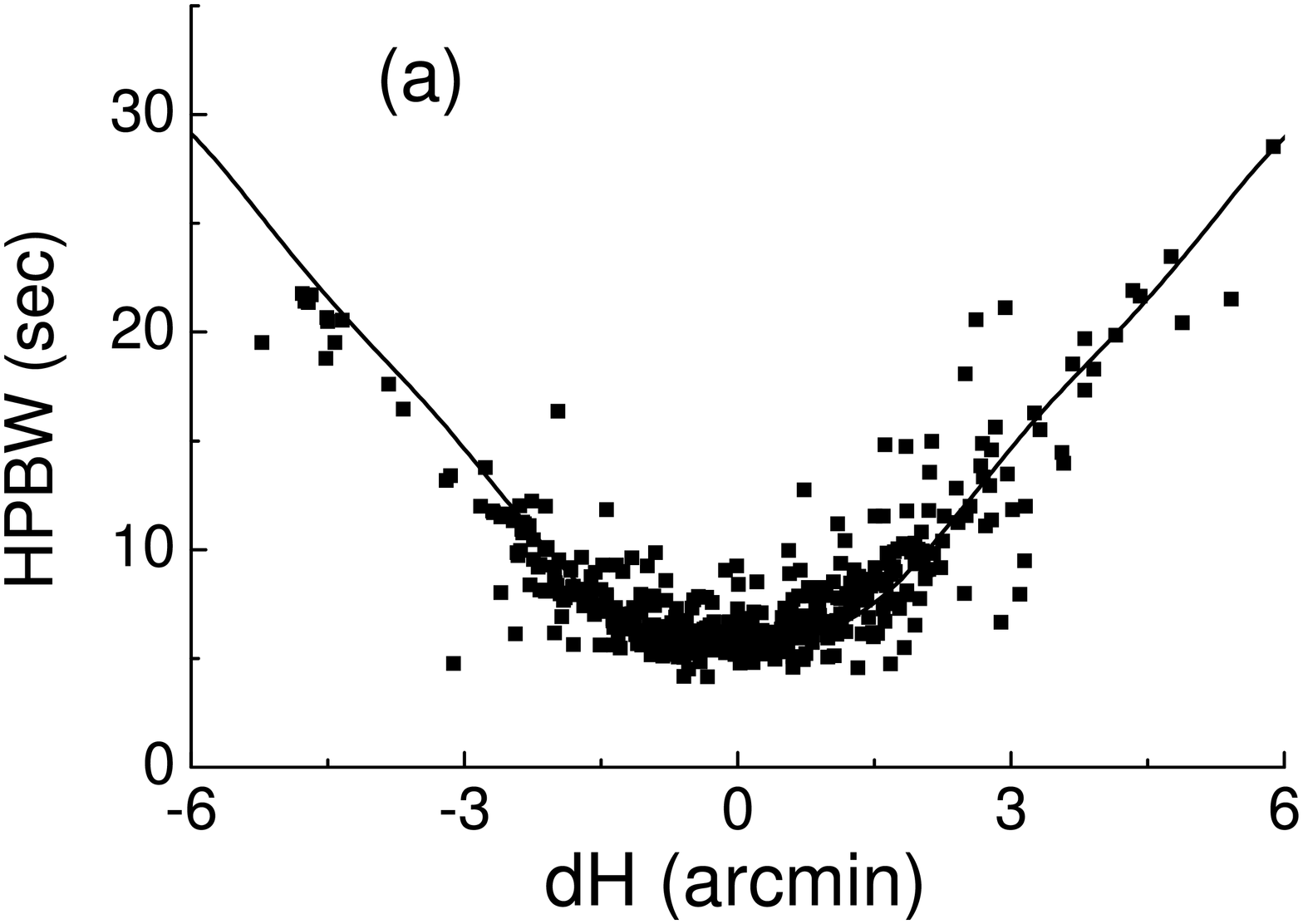}
\includegraphics[angle=-90,width=0.33\textwidth,clip]{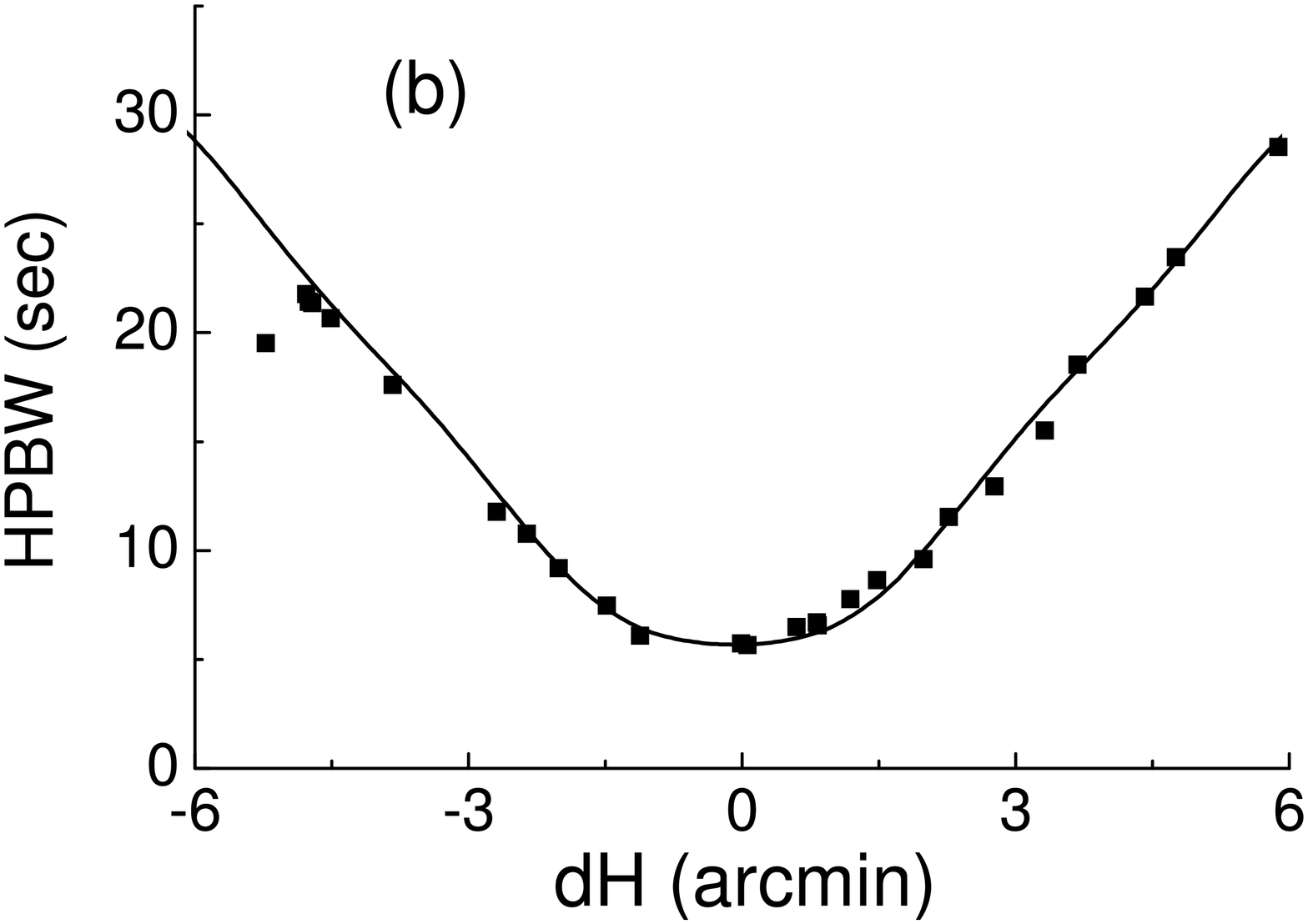}
}
}
\caption{
The $dH$ dependences of the halfwidth of the vertical PB constructed using  the sources
identified on simulated scans (filled squares): for the entire sample of sources ($S \ge
2.7$~mJy) (a) and for sources with fluxes $S \ge 200$~mJy (b). The simulated scans were
obtained from   $1\degr\times 1\degr$ NVSS images. The solid lines show the vertical
section of the computed PB at 7.6~cm after its convolving in each horizontal section with a
Gaussian of size $\theta_{0.5} = 45^{''}$.
}
\label{fig13:Majorova_n}
\end{figure*}

\begin{figure*}[tbp]
\centerline{
\hbox{
\includegraphics[angle=-90,width=0.30\textwidth,clip]{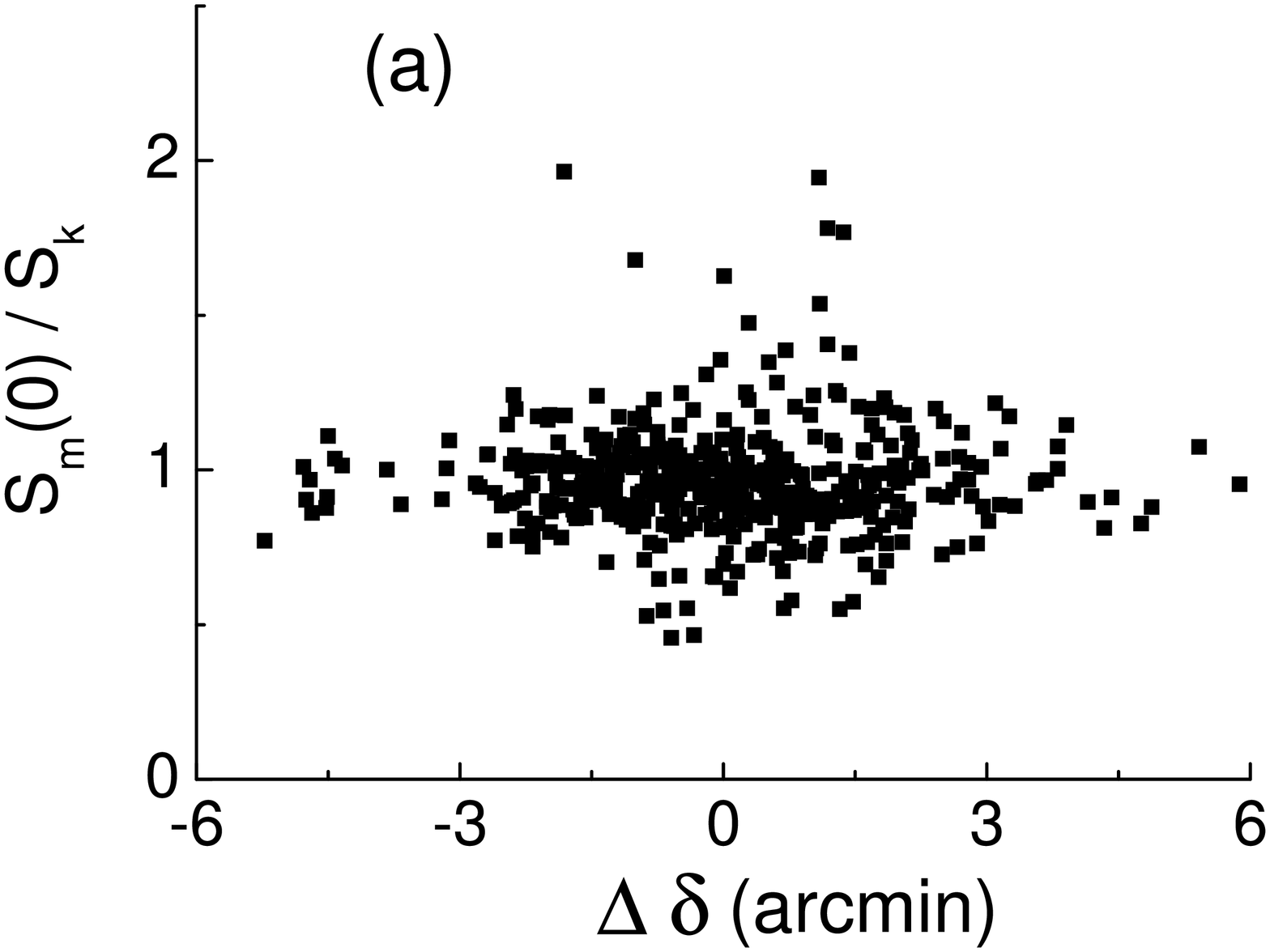}
\includegraphics[angle=-90,width=0.30\textwidth,clip]{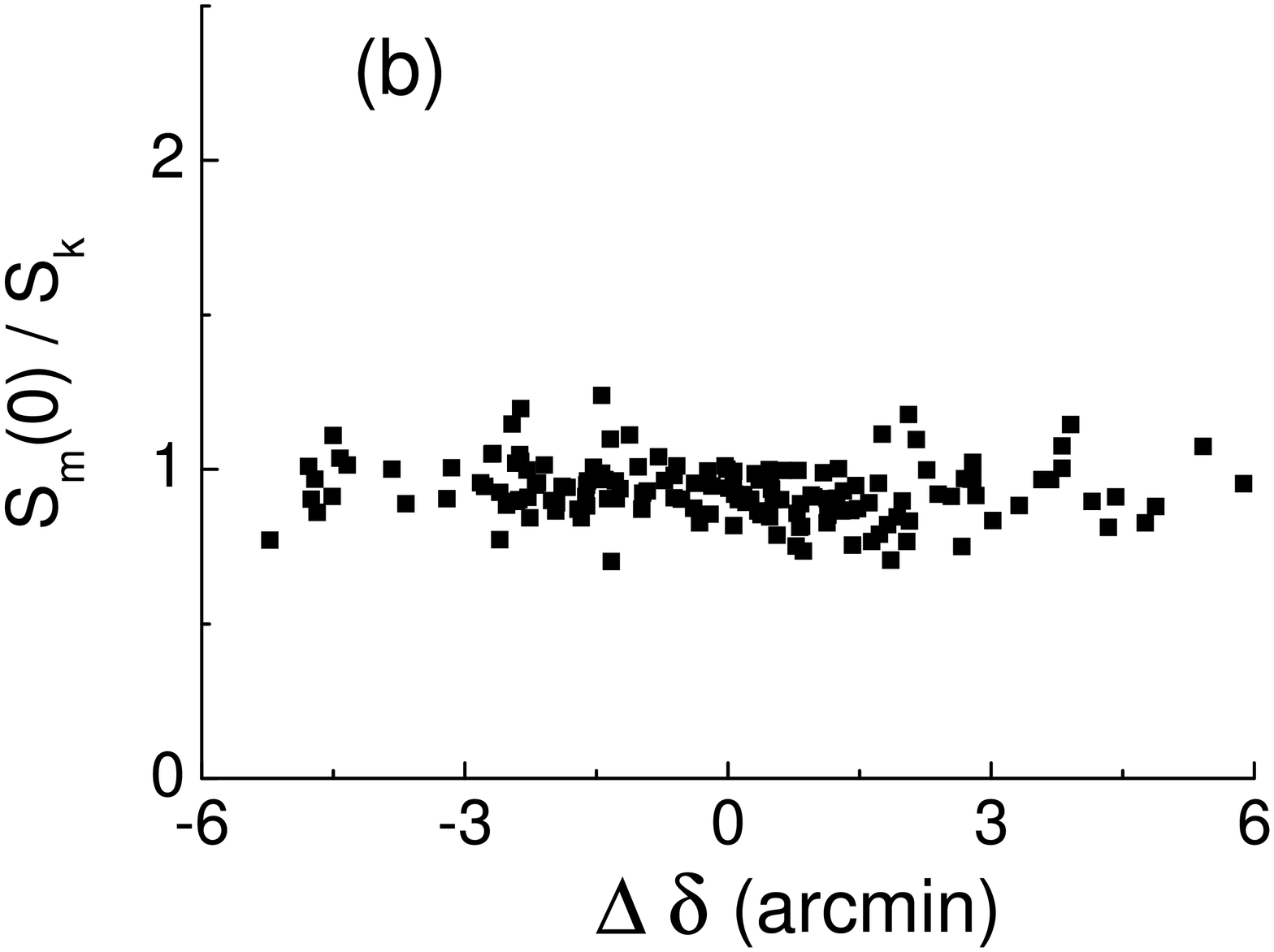}
\includegraphics[angle=-90,width=0.30\textwidth,clip]{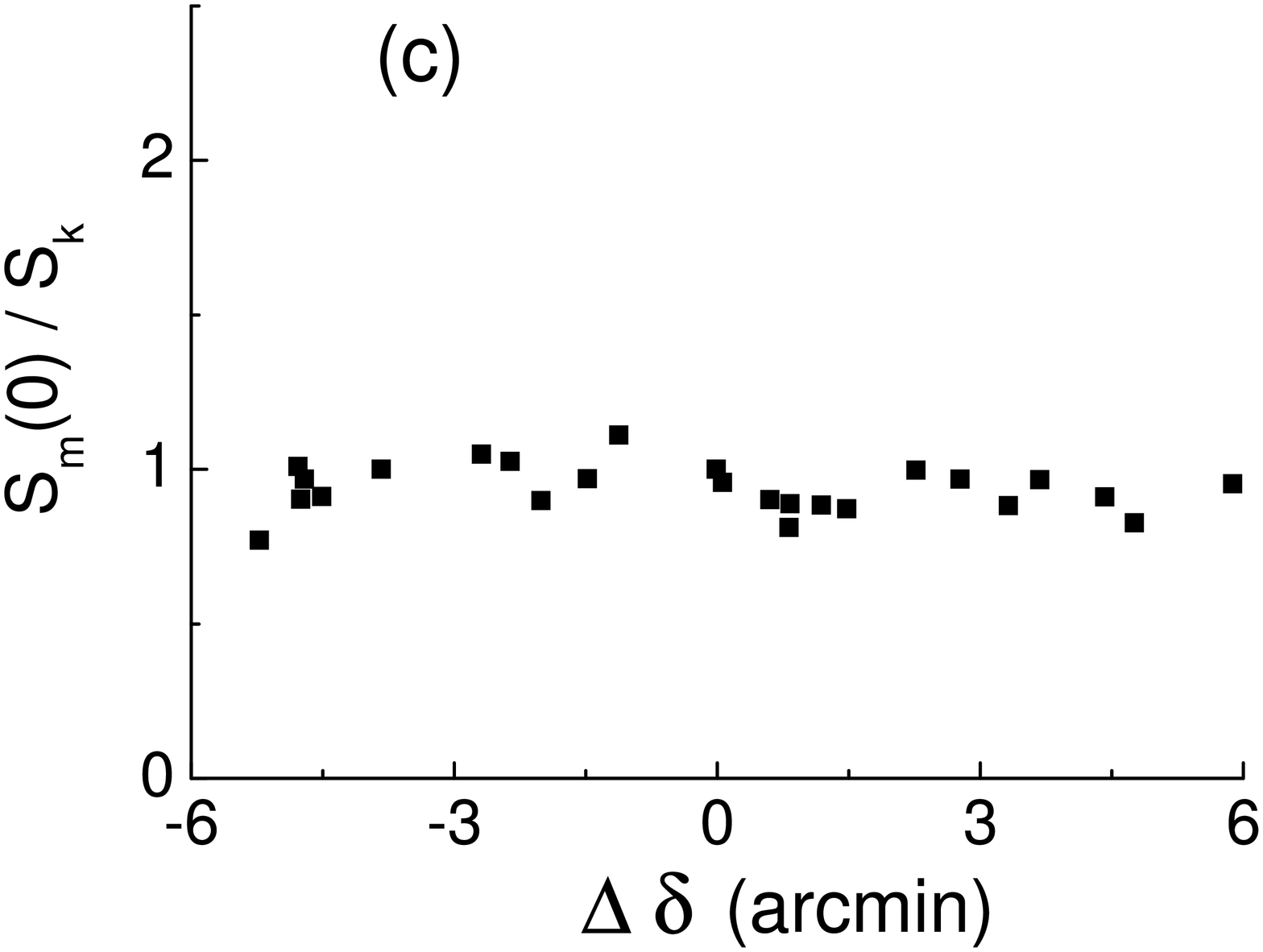}
}
}
\caption{
The $\Delta \delta$ dependences of $S_{m}(0)/S_{k}$ constructed using the entire sample of
sources identified on simulated scans ($S \ge 2.7$~mJy) (a) and using sources with fluxes $S \ge
30$~mJy (b) and $S \ge 200$~mJy (c). The simulated scans were obtained from   $1\degr\times 1\degr$ NVSS images.
}
\label{fig14:Majorova_n}
\end{figure*}

\begin{figure*}[tbp]
\centerline{
\hbox{
\includegraphics[angle=-90,width=0.30\textwidth,clip]{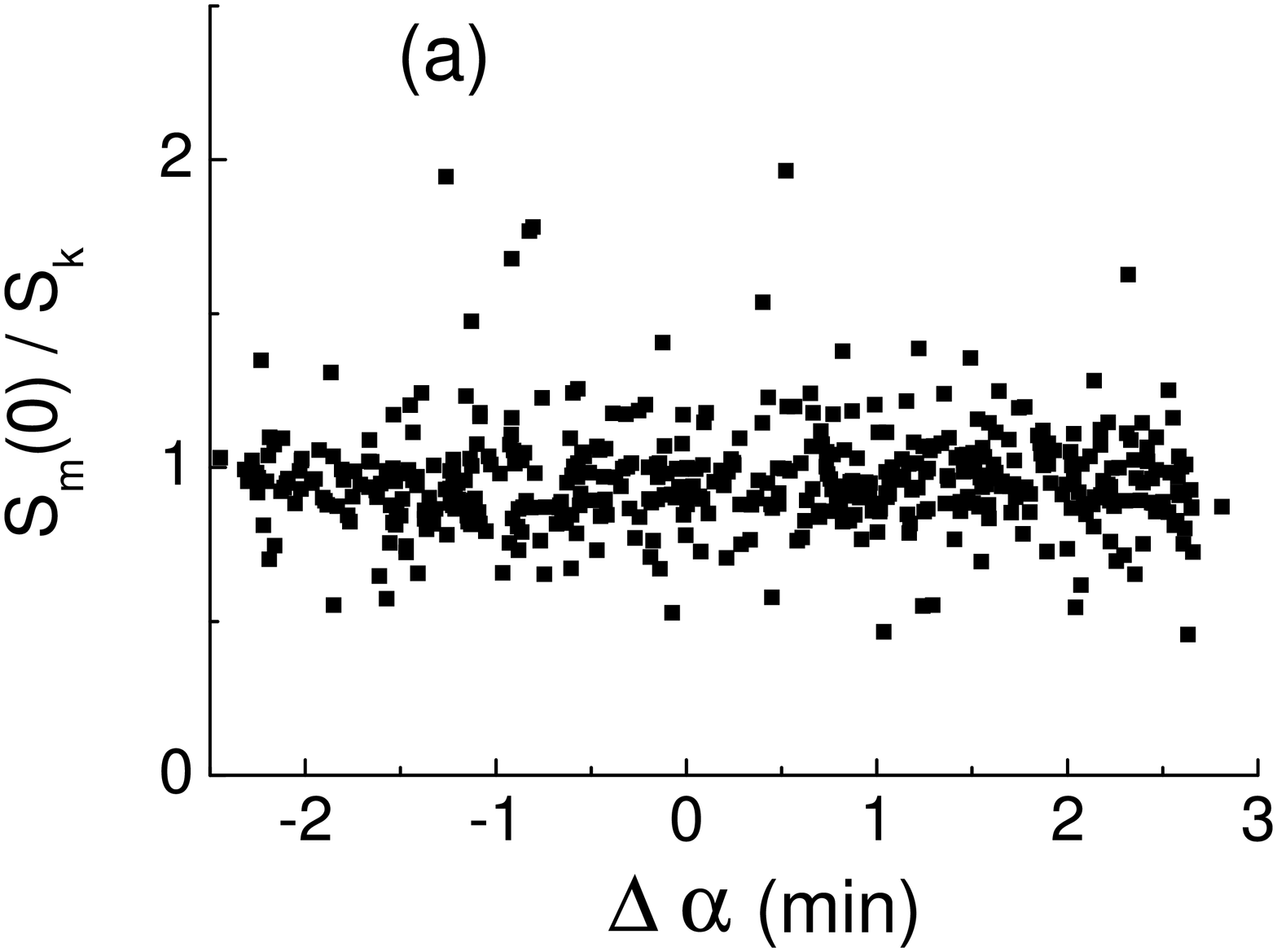}
\includegraphics[angle=-90,width=0.30\textwidth,clip]{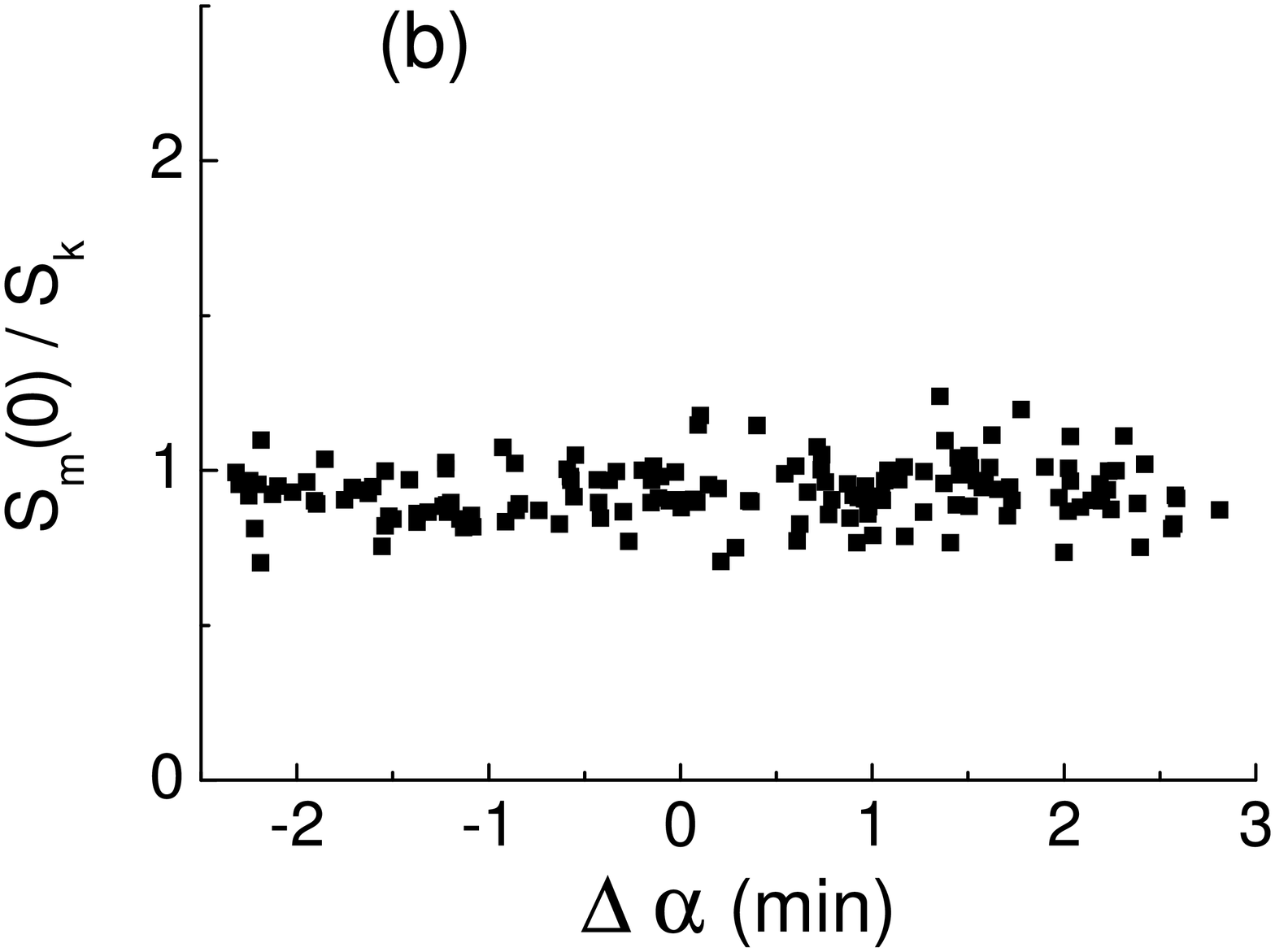}
\includegraphics[angle=-90,width=0.30\textwidth,clip]{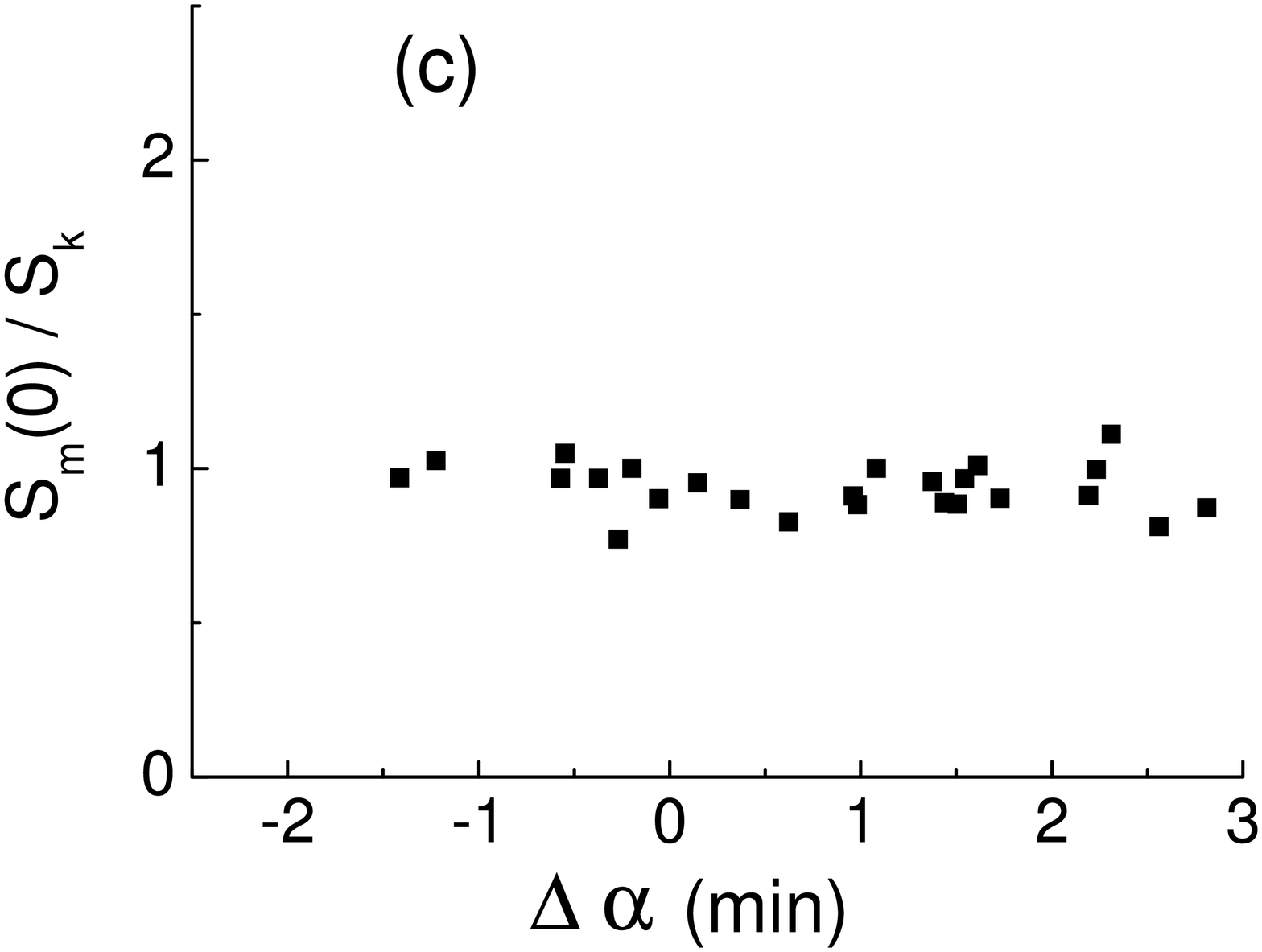}
}
}
\caption{
The $\Delta \alpha$ dependences of $S_{m}(0)/S_{k}$ constructed using the entire sample of
sources identified on simulated scans ($S \ge 2.7$~mJy) (a) and using sources with fluxes $S \ge
30$~mJy (b) and $S \ge 200$~mJy (c). The simulated scans were obtained from   $1\degr\times
1\degr$ NVSS images. }
\label{fig15:Majorova_n}
\end{figure*}

We use Gaussian analysis to identify the sources of the NVSS
catalog on simulated scans; determine the amplitude (in relative
and absolute units) and halfwidth of their signals. We then use
the results obtained to construct the same dependences as those
constructed for the sources identified on 20-minute simulated
scans, namely: the $\Delta \delta$ and $\Delta \alpha$
dependences of $S_{m}(0)/S_{k}$, the $dH$ dependence of HPBW, and
the vertical PB of the radio telescope. We show the results in
Figs.~\ref{fig12:Majorova_n}, ~\ref{fig13:Majorova_n},
~\ref{fig14:Majorova_n}, and ~\ref{fig15:Majorova_n}.

As is evident from the figures, the vertical PB and the $dH$
dependence of HPBW constructed using five-minute simulated scans
agrees well with the computed curves; the $\Delta \delta$
dependences of $S_{m}(0)/S_{k}$ have zero slopes, and the $\Delta
\alpha$ dependences of $S_{m}(0)/S_{k}$ show a uniform scatter of
data points along the entire scan length. This fact indicates
that simulated scans contain no systematic errors and that they
can be used to obtain qualitative estimates and not just to
detect and identify the sources of the NVSS catalog on real scans.

\section{ESTIMATION OF THE ACCURACY OF THE INFERRED SOURCE FLUXES}

Identification of NVSS sources on simulated scans obtained by convolving NVSS images with the
power beam pattern of RATAN-600 and subsequent reduction of their fluxes to the central
section of the survey must yield the fluxes given in the NVSS catalog. However, the
dependences shown in Figs.~\ref{fig14:Majorova_n}\,(a,b,c) and
\ref{fig15:Majorova_n}\,(a,b,c) demonstrate evident scatter of data points about
$S_{m}(0)/S_{k}=1.0$, and this scatter increases with decreasing source flux.

\begin{figure*}[tbp]
\hbox{
\centerline{
\includegraphics[angle=-90,width=0.35\textwidth,clip]{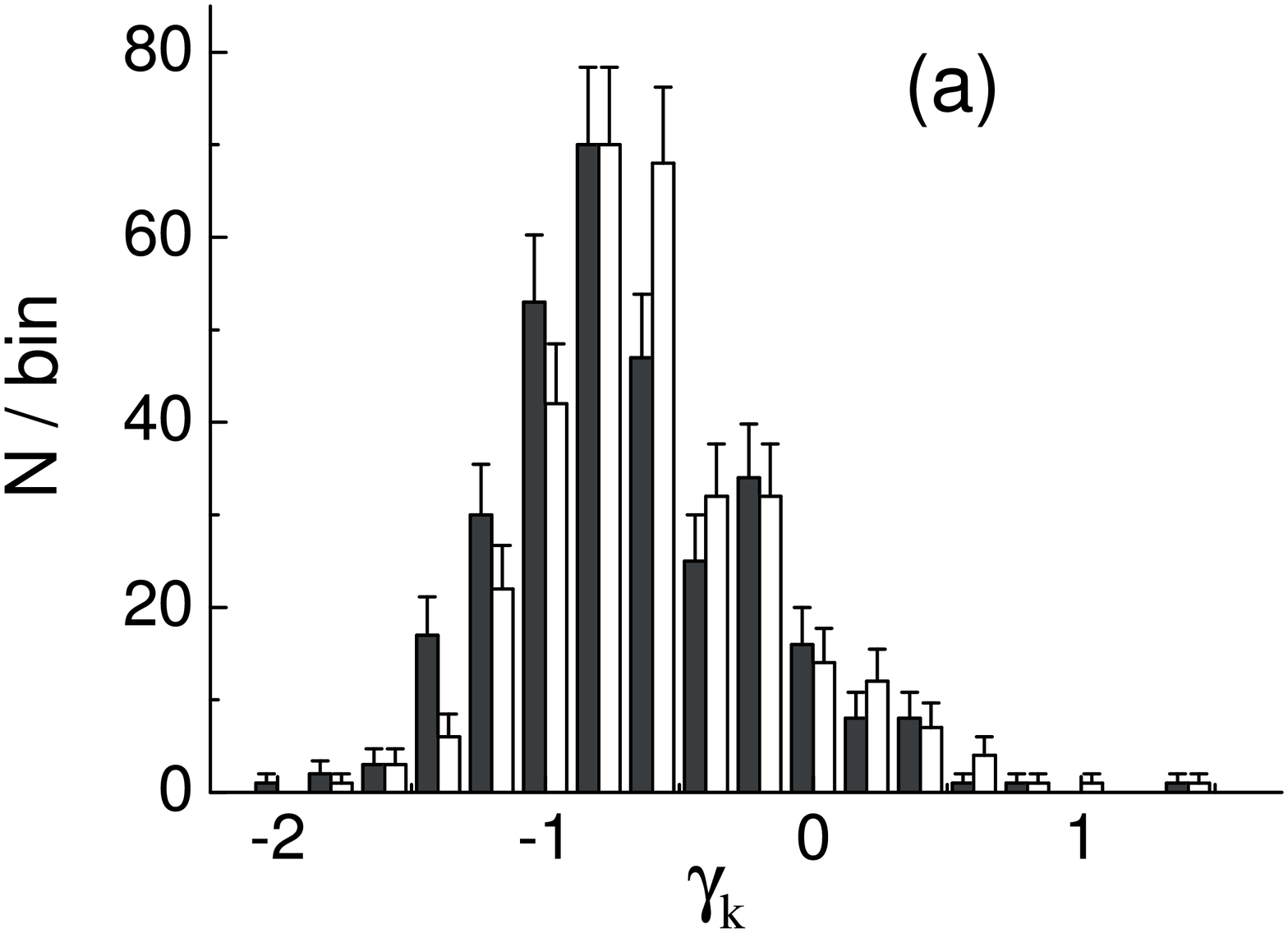}
\includegraphics[angle=-90,width=0.35\textwidth,clip]{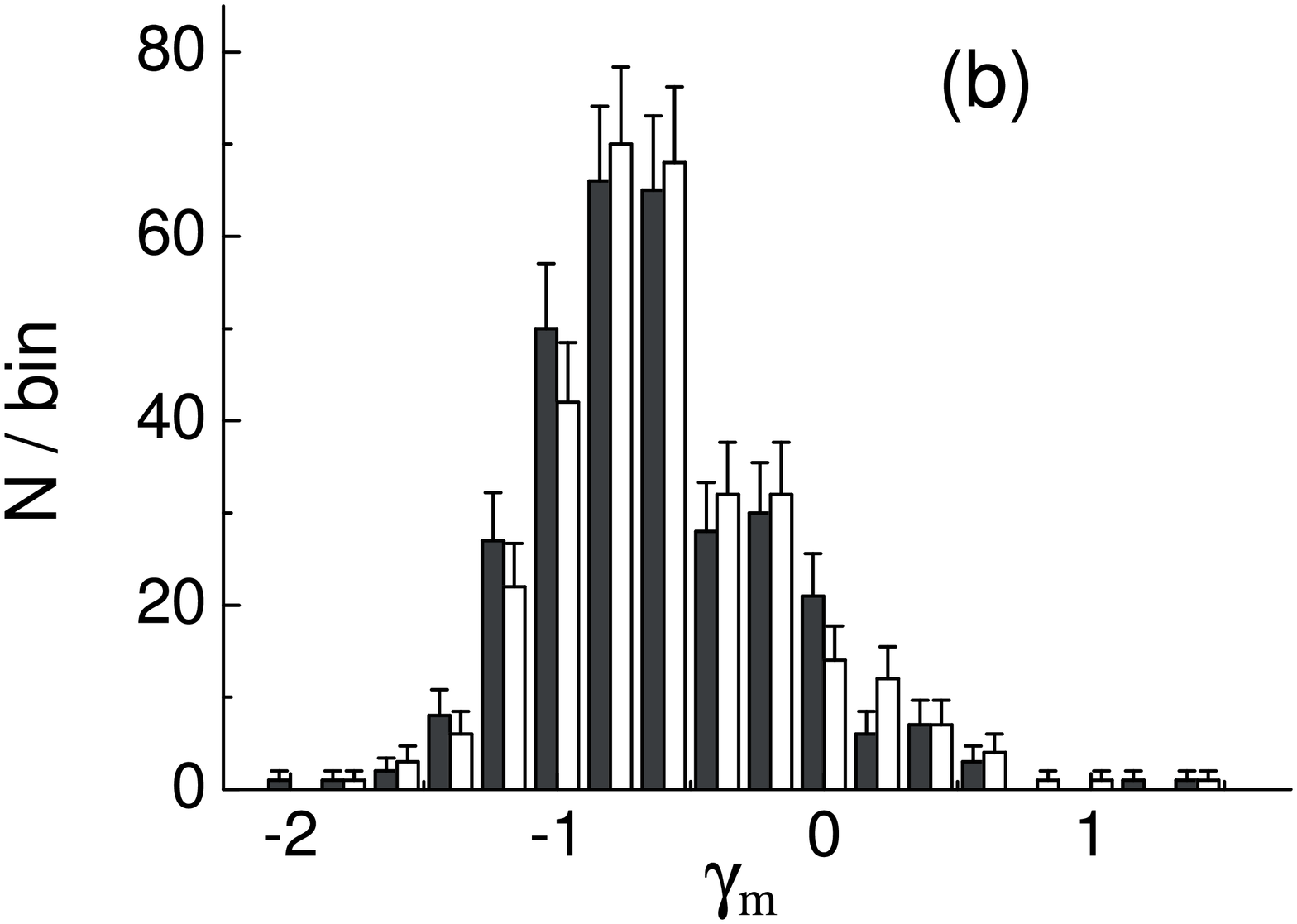}
}
}
\caption{
Histograms of spectral indices obtained using the fluxes listed in the NVSS
catalog, $\gamma = ln(S_{7.6} / S_{21}^{k})$ (a) (open bars), and using the fluxes
inferred from simulated scans based on NVSS images, $\gamma = ln(S_{7.6} / S_{21}^{m})$ (b)
(open bars). The histogram of spectral indices based on the data of the RZF
catalog (filled bars) [\cite{p1:Majorova_n}]. The histograms are based on a total sample
of 317 sources with  allowance for errors of inferred spectral indices.
}
\label{fig16:Majorova_n}
\end{figure*}

This is because as source fluxes decrease, source signals and the noise from
background sources become close in magnitude, thereby making extraction of sources
difficult both on simulated and real scans. Cases occur where sources blend together (are
superimposed on each other). In some cases even strong sources are identified with
insufficient confidence if they have close right ascensions, but are located at different
distances from the central section of the survey. All the above effects result in the
distortion of signals from sources and even in the appearance of false sources.

The errors of the inferred fluxes can be estimated using the resulting $\Delta\alpha$
dependences of $S_{m}(0)/S_{k}$. The standard error $\sigma_{S}^{m}$  is equal to the standard
deviation of $S_{m}(0)/S_{k}$ from its mean value averaged over the entire sample of the
sources identified.

This error gives the lower limit for the error of the
determination of fluxes of sources identified on real scans at
$\lambda7.6$~cm. Recall that the error of the inferred fluxes on
real scans also includes the error of interpolation
(extrapolation) of fluxes to the wavelength of  7.6~cm. Other
error components should be of similar magnitude, because we use
the same method to identify sources and use the same instrumental
function of the radio telescope to convert the source fluxes to
their values in the central section of the survey.

The standard error of the determination of the fluxes of the
sources identified on simulated scans is: $\sigma_{S}^{m}
=0.20\pm0.01$---for the entire sample of the sources
considered~~  with~~ fluxes $S > 2.7$~mJy; $\sigma_{S}^{m}
=0.10\pm0.01$,~~ for~~ sources with $S > 30$~mJy and \mbox{$\sigma_{S}^{m}
=0.08\pm0.01$,} for sources with $S > 200$~mJy.  (Here we give the
source fluxes at $\lambda21$~cm).

Let us now compare the errors of  $\sigma_{S}^{m}$ with the errors $\sigma_{S}^{r}$ of the
determination of the fluxes of sources identified in real records. The errors
$\sigma_{S}^{r}$  determined by Bursov et al. [\cite{p1:Majorova_n}] and Majorova and
Bursov [\cite{m3:Majorova_n}] based on the entire sample of sources considered exceed
$\sigma_{S}^{m}$ by $20\div25\%$ ($\sigma_{S}^{r} = 0.24\pm0.03$ [\cite{m3:Majorova_n}],
$\sigma_{S}^{r} = 0.25$ [\cite{p1:Majorova_n}]). For strong sources the error $\sigma_{S}^{r}$
 was equal to   $0.15\pm0.03$
[\cite{p1:Majorova_n,m3:Majorova_n}].

As we pointed out above, such discrepancies between
$\sigma_{S}^{m}$ and $\sigma_{S}^{r}$ are due primarily to errors
of interpolation (extrapolation) of fluxes from $\lambda21$~cm to
$\lambda7.6$~cm. However, a certain deviation of the real PB from
the computed PB during the survey may also contribute to the
above discrepancy.

\section{DETERMINATION OF THE SPECTRAL INDICES OF THE SOURCES}

The spectral index of a source can be computed as the logarithm
of the  $S_{7.6} / S_{21}^{k}$ ratio or the logarithm of the
$S_{7.6} / S_{21}^{m}$ ratio, where $S_{7.6}$ is the flux of the
source as inferred from the real averaged record of the RZF
survey at $\lambda7.6$~cm; $S_{21}^{k}$ is the $\lambda21$~cm
flux of this source according to the data of the NVSS catalog,
and $S_{21}^{m}$ is the source flux in mJy as inferred from the
simulated scan.

Given that we use the same technique to identify sources on real
and simulated scans, and given the high degree of correlation
between these scans, one may assume that the errors of the
extraction of the same source should be the same for both
scans. That is, we should either overestimate or underestimate the
flux of a source when identifying it simultaneously in the
simulated and real records. Thus the spectral indices computed by
the formula $\gamma_{m} = ln(S_{7.6} / {S_{21}}^{m})$ may prove
to be even more accurate than those inferred from the data of the
NVSS catalog ($\gamma_{k} = ln(S_{7.6} / S_{21}^{k})$). This is
true if the computed and real power beams of the radio
telescope agree well with each other.

To compare the spectral indices of the sources obtained using
different methods, we identify on the averaged 7.6-cm records of
the RZF survey the sources of the NVSS catalog
within the  $\pm6'$ band of the central section of the RZF survey.
 We identified
a total of about 500  NVSS-catalog sources on the 24-hour record
of the sky scan, determine the spectral indices $\gamma_{k}$ and
$\gamma_{m}$ for each of these sources, and compare them to the
spectral indices obtained by Bursov et al.~[\cite{p1:Majorova_n}].
For a more correct comparison, we used only the sample of 317
sources in common with the RFZ catalog.

The open bars in Fig.~\ref{fig16:Majorova_n} show the histograms
of the distribution of spectral indices obtained using the fluxes
listed in the NVSS catalog, $\gamma_{k} = ln(S_{7.6} /
S_{21}^{k})$ (a) and the distribution of spectral indices computed
using the fluxes obtained from simulated scans $\gamma_{m} =
ln(S_{7.6} / S_{21}^{m})$ (b). The histograms are based on the
sample of 317 sources that we identified on simulated and real
scans. The filled bars in the same figure show the histograms of
the spectral indices of the same sample of sources based on the
data of the RZF catalog [\cite{p1:Majorova_n}]. These histograms
were constructed with allowance made for the errors of the
determination of spectral indices.

As is evident from the figures shown, the histogram of the
spectral indices $\gamma_{m}$ computed using the fluxes inferred
from simulated scans proves to be closer to the distribution of
spectral indices according to the data of the RZF catalog for the
source sample considered. Given the confidence interval, they can
be considered  to be virtually identical.

It follows from this that given the real records and the
corresponding simulated scans obtained using NVSS images, one can
infer the spectral indices of the sources from the ratio of their
signals in the scans considered. The real and simulated scans
must be normalized prior to that. Such a procedure is very easy and
convenient to perform for sources whose spectral indices are
unknown.

\section{ESTIMATE OF RESIDUAL NOISE ON REAL RECORDS USING
SIMULATED SCANS}

We use the simulated scans based on NVSS images to clean the real
records from discrete sources. To this end, we ``join'' the
normalized five-minute scans into a single 24-hour scan and
convert it into a 7.6-cm wavelength scan in accordance with the
average spectral index.

As we pointed out above, in this method it is hardly possible to
correct the signal from each source in accordance with its own
spectral index even if the spectral index is known, and therefore
we used for our conversion the spectral indices close to those of
the sources with normal nonthermal spectra, which lie in the
$\gamma = 0.78\pm0.02$ interval. We chose the spectral index so as
to minimize the standard error of the deviation of the simulated
record corrected for the spectral index  from the real record. We
used trimmed records $ 0^{h} \le \alpha < 20^{h}$ (with the
Galactic plane excluded) to estimate the magnitude of residual
noise.

The standard error of residual noise at 7.6~cm was equal to
$\sigma = 7.6\pm0.9$~mJy. This residual noise includes the noise
due to unresolved sources; radiometer noise; atmospheric noise,
and residual signals from the sources whose spectral indices
differ from the average spectral index used to convert the source
fluxes.

Subtraction of the background inferred with a $ 30^{s}$ smoothing
window (hereafter referred to as the thirty-second background)
partially eliminates atmospheric noise and the noise from the
sources located outside the $\Delta\delta = 41\degr30'42''\pm6'$
band. This procedure  reduces  $\sigma$ of residual noise down to
$4.6\pm0.6$~mJy. Note, for comparison, that the magnitude of
residual noise after we subtract from the real averaged record
the simulated scan obtained from the data of the NVSS catalog and
then the thirty-second noise, is equal to  $\sigma =
5.9\pm0.6$~mJy. This means that cleaning real records using
simulated scans obtained by applying two different methods yields
similar results.

\begin{figure}[tbp]
\centerline{
\vbox{
\hbox{
\includegraphics[angle=-90,width=0.43\textwidth,clip]{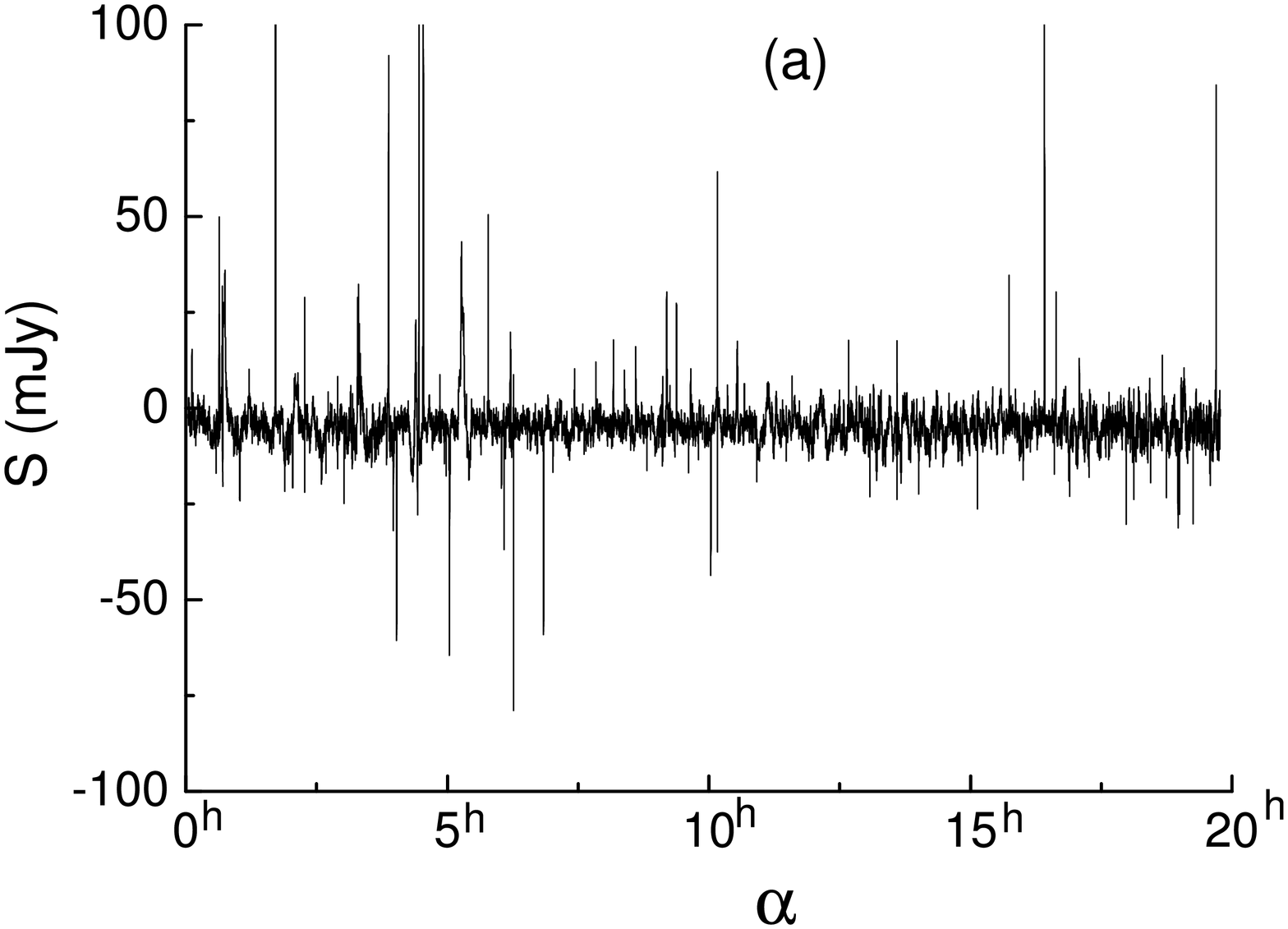}
}
\hbox{
\includegraphics[angle=-90,width=0.43\textwidth,clip]{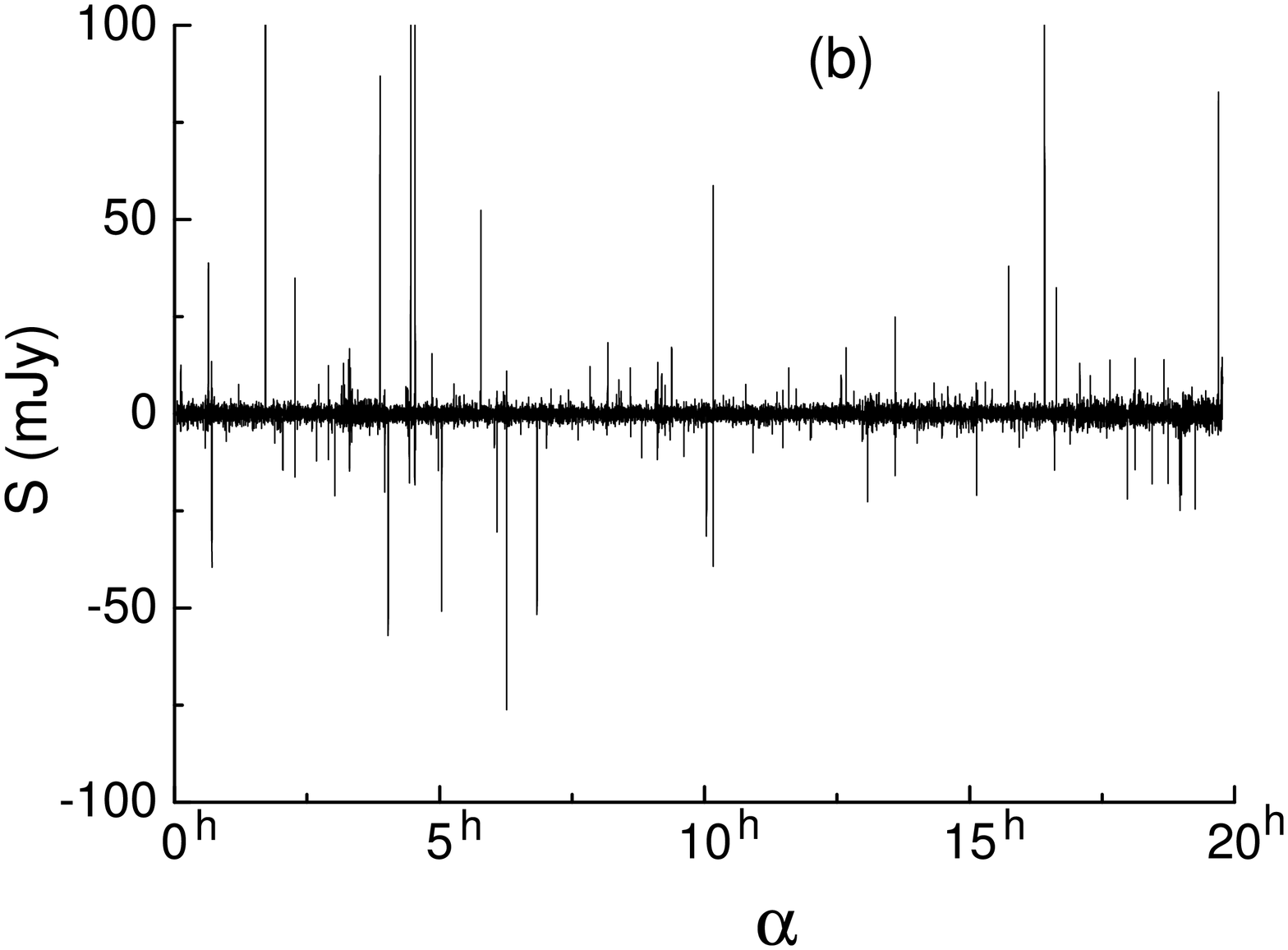}
}
\hbox{
\includegraphics[angle=-90,width=0.43\textwidth,clip]{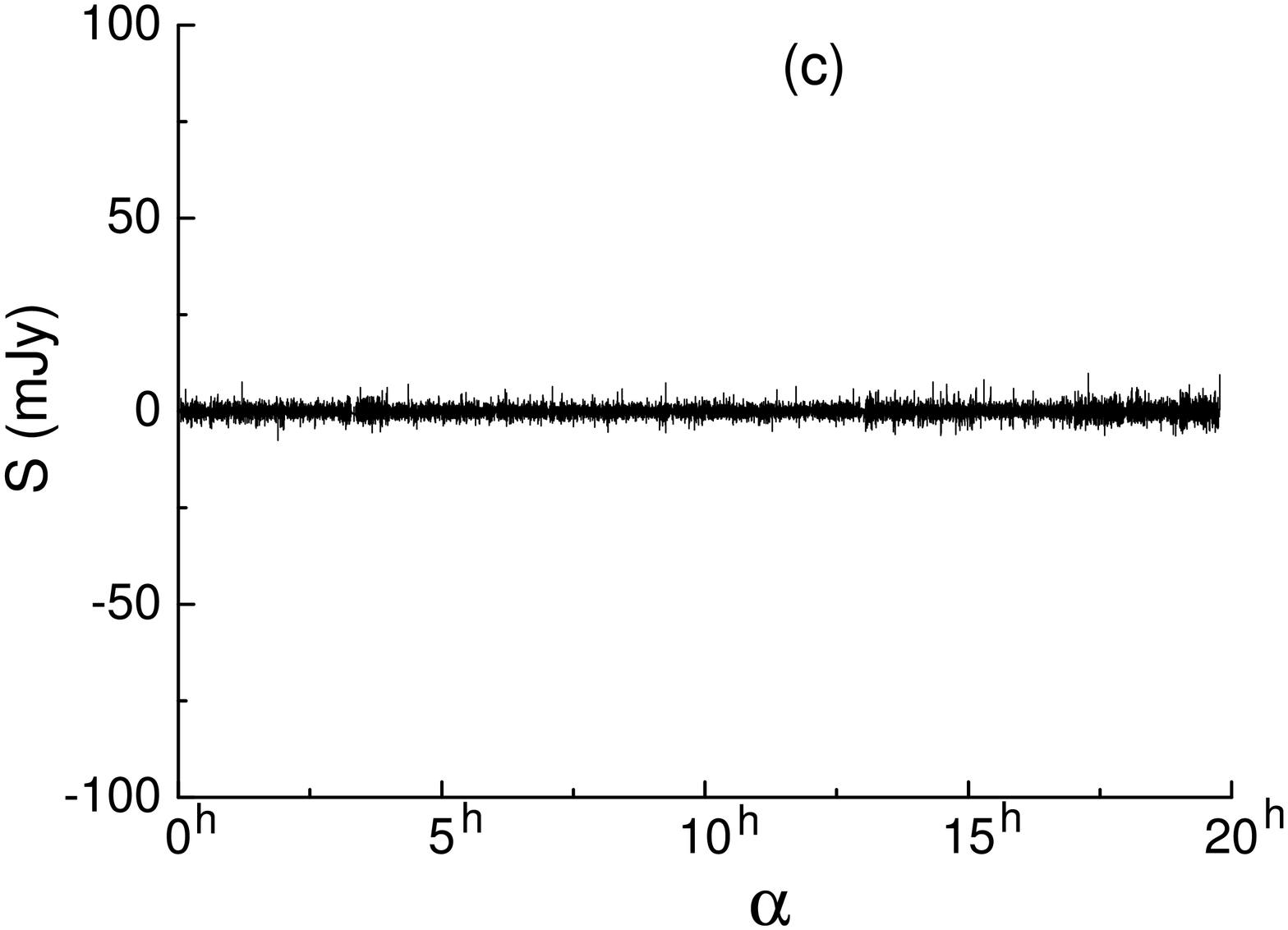}
}
}
}
\caption{
Residual noise on a real averaged record of the RFZ survey after subtracting the simulated
scan obtained using NVSS images (a); residual noise after subtracting the background obtained
with a $30^{s}$ smoothing window from the difference of the real and simulated scans (b), and
residual noise after subtracting the sources whose spectral indices differ from the average
spectral index (Ó).
}
\label{fig17:Majorova_n}
\end{figure}

Figure~\ref{fig17:Majorova_n}(a) shows the residual noise on a real averaged record of the
RZF survey after subtracting from it the simulated scan based on NVSS images and corrected for
the average spectral index. Figure~\ref{fig17:Majorova_n}(b) shows the residual noise obtained
after further subtraction of the background computed with a $ 30^{s}$ smoothing window from the
difference of the real and simulated scans.

As is evident from the figures, the records still contain signals from the sources whose
spectral indices differ from the average spectral index. After eliminating such signals
(they can be easily seen in the records) the standard deviation $\sigma$ of the residual noise
becomes equal to about 1~mJy. Figure~\ref{fig17:Majorova_n}(c) shows the record of the
residual noise. Note that the standard error of the same portion of the uncleaned real record
($ 0^{h} \le \alpha < 20^{h}$) is equal to $19\pm3$~mJy. We did not take the 3C84 signal into
account when estimating $\sigma$. Thus cleaning of records using simulated scans reduces the
residual noise of a real record by more than one order of magnitude.

Note that records can be cleaned even without the use of simulated scans, simply by ``cutting''
the sources on records in accordance with their coordinates. However, such a cleaning procedure
is very time consuming and sometimes ambiguous, because the transit curves obtained in
RATAN-600 surveys are superpositions of source signals convolved  with the PB at different
declinations. The proposed method allows easy elimination of most of the point sources within
the given declination band and with fluxes greater than $3$~mJy ($> 3\sigma$). More refined
methods are required for a more in-depth cleaning, and we do not describe them here, because
they are a subject of a separate paper.

\section{CONCLUSIONS}

We propose and develop a method for the simulation of a deep multi-frequency zenith-field
sky survey by convolving  NVSS images with two-dimensional power beams of RATAN-600.

We perform simulations at  1.0, 2.7, 3.9, 7.6, 13, 31, and 49~cm in the central band of the
survey. We obtained 24-hour simulated scans of the sky transit across the power beam pattern
 of the telescope at each of the wavelengths listed above and compared the simulated
scans with real records and with the simulated scans based on the data of the NVSS
catalog [\cite{b1:Majorova_n}].

We analyze the effect of the size of NVSS-image areas on the results of simulations
at 7.6~cm. We show that simulated scans based on $4\degr\times 4\degr$ ($20^{m}$ in right
ascension) NVSS image areas are highly correlated (with a correlation coefficient of $\sim
90\%$) with real scans and can therefore be used to identify and extract weak sources in the
records of the RZF survey. However, for quantitative estimates the size of NVSS image areas
should be reduced down to  $1\degr\times 1\degr$ ($5^{m}$ in right ascension).

We obtained a total of 288 five-minute simulated scans of sky transits at the wavelength of
7.6~cm by convolving  $1\degr\times 1\degr$ NVSS images with the PB of RATAN-600, and then
``joined'' them into a single 24-hour scan. We identified a total of about 500 NVSS-catalog
sources simultaneously in simulated scans and real averaged records of the RZF survey
within the  $\pm6'$ band of the central section of the survey
and  estimated their fluxes.

The data obtained allowed us to estimate the accuracy of the determination of source fluxes
on simulated scans. The standard error of the determination of source fluxes on simulated scans
is  $\sigma_{S}^{m} =$ \mbox{$0.20\pm0.01$} for the entire sample of sources considered
\mbox{($S > 2.7$~mJy).} This error is $20\div25\%$ lower than the standard error for real records
[\cite{p1:Majorova_n,m3:Majorova_n}]. The standard error is  $\sigma_{S}^{m} =
0.10\pm0.01$ for sources with fluxes exceeding 30~mJy. The standard errors $\sigma_{S}^{m}$
obtained can be viewed as a lower limit for the errors of the determination of source fluxes on
real scans at $\lambda7.6$~cm.

We derived the distributions of spectral indices of the sources by identifying sources on real
and simulated scans. To compute the spectral indices, we used the source fluxes listed in the
NVSS catalog and the source fluxes inferred from real and simulated scans. We compared the
distributions of spectral indices obtained using different methods with the distribution of
spectral indices based on the data of the RZF catalog for the common sample of 317 sources.
We show that the distribution of spectral indices computed as the logarithm of the ratio of
the fluxes of sources identified on real and simulated scans agrees within the quoted errors
with the distribution of spectral indices inferred from the data of the RZF survey of the same
sample of sources.

We use the simulated scans obtained to clean the real 20-hour record of the RZF survey.
After the subtraction of the thirty-second noise and elimination of residual signals of the
sources whose spectral indices differ from the mean spectral indices the standard deviation
$\sigma$ of residual noise was equal to $\sim 1$~mJy at 7.6~cm. This procedure reduced the
noise of the initial record of the RZF survey by more than one order of magnitude.

\begin{acknowledgements}
I am grateful to Yu.~N.~Parijskij for his support of the work and
discussion of the results, N.~N.~Bursov for sharing the data of
the RZF survey and simulated scans based on the NVSS catalog.

This work was supported in part by the Russian Foundation for Basic Research (grant
no.~05-02-17521) and the Program for the Support of Leading Scientific Schools of the
President of the Russian Academy of Sciences (the ``School of S.~E.~Khaikin'').
\end{acknowledgements}


\begin{thebibliography}{99}

\bibitem{p2:Majorova_n}[1]
Yu.N.Parijskij and D.V.Korol'kov,
{\it Itogi Nauki i Tekhniki. Astrofiz. Kosmicheskaya Fiz. Edited by R.A.Sunyaev. Seriya
Astronomiya} (VINITI, Goskomitet po Nauke i Tekhnike, Moscow, 1986)  \textbf {31},  73 (1986).

\bibitem{p3:Majorova_n}[2]
Yu.~N.~Parijskij,  N.~N.~Bursov,  N.~M.~Lipovka,  et al., Astron.~and Astrophys. Suppl.
 \textbf{87}, 1 (1991).

\bibitem{p4:Majorova_n}[3]
Yu.~N.~Parijskij,  N.~N.~Bursov,  N.~M.~Lipovka,  et al., Astron.~and Astrophys. Suppl.
 \textbf{96}, 583 (1992).

\bibitem{a1:Majorova_n}[4]
{\it Catalog of Radio Sources of the Zelenchuk Sky Survey in the Declination Interval
$0\degr - 14\degr$}, Edited by M.~G.~Larionov (Izdatel'stvo Moskovskogo Universiteta, 1989).

\bibitem{a2:Majorova_n}[5]
V.~R.~Amirkhanyan, A.~G.~Gorshkov,  and A.~V.~Ipatov, Soobshch. Spets. Astrofiz. Obs. {\bf
58}, 41 (1988).

\bibitem{a3:Majorova_n}[6]
V.~R.~Amirkhanyan, A.~G.~Gorshkov, A.~A.~Kapustin, et al., Pis'ma Astron. Zhurn. {\bf 18},  396
(1992).

\bibitem{v1:Majorova_n}[7]
M.~G.~Mingaliev, O.~V.~Verkhodanov, and A.~R.~Khabrakhmanov, Pis'ma Astron. Zhurn. {\bf 17},
787 (1991).

\bibitem{p1:Majorova_n}[8]
N.~N.~Bursov, Yu.~N.~Parijskij, E.~K.~Majorova, et al.,
Astron. Zhurn. {\bf 84},  227 (2007).

\bibitem{co1:Majorova_n}[9]
J.~J.~Condon, W.~D.~Cotton, E.~W.~Greisen, et al.,
Astronom. J.  {\bf115}, 1693 (1998)

\bibitem{fr:Majorova_n}[10]
R.~L.~White, R.~H.~Becker, D.~J.~Helfand, and M.~D.~Gregg,
Astrophys.J. {\bf 475}, 479 (1997).

\bibitem{m1:Majorova_n}[11]
E.~K.~Majorova,
Bull. Spec. Astrophys. Obs. {\bf 53}, 78 (2002).

\bibitem{m2:Majorova_n}[12]
E.~K.~Majorova and S.~A.~Trushkin,
Bull. Spec. Astrophys. Obs. {\bf 54}, 89 (2002).

\bibitem{b1:Majorova_n}[13]
N.~N.~Bursov and E.K.Majorova, in
{\it Abstracts of Papers. Russian Conference Dedicated to the Memory of A.~A.~Pistel'kors
``Radio telescopes RT-2002'', Pushchino, October 9--11, 2002}, 26, (2002).

\bibitem{na1:Majorova_n}[14]
J.~J.~Condon, W.~D.~Cotton, E.~W.~Greisen, et al., \\
http://www.cv.nrao.edu/nvss/,\\
http://ftp.cv.nrao.edu/fits/os-support/unix/xfitsview/.

\bibitem{na2:Majorova_n}[15]
SkyView,
http://skyview.gsfc.nasa.gov/cgi-bin/skvbasic.pl.

\bibitem{b2:Majorova_n}[16]
N.~N.~Bursov, Candidate's Dissertation  in Physics and Mathematics (Special Astrophysical Observatory,
Russian Academy of Sciences, Nizhnii Arkhyz, 2003).

\bibitem{m3:Majorova_n}[17]
E.~K.~Majorova and N.~N.~Bursov,
Astrophys. Bull. {\bf 62}, 398 (2007).

\end{thebibliography}
\end{document}